\documentclass[onecolumn,journal]{IEEEtran}

\usepackage{cite}

\ifCLASSINFOpdf
\else
   \usepackage[dvipdfmx]{graphicx}
\fi
\usepackage[cmex10]{amsmath}
\usepackage{amssymb}
\usepackage{balance}
\newtheorem{property}{Property}
\newtheorem{theorem}{Theorem}
\newtheorem{lemma}{Lemma}
\newtheorem{definition}{Definition}
\newtheorem{assumption}{Assumption}
\newtheorem{remark}{Remark}

\newtheorem{proposition}{Proposition}

\begin{document}
%
\title{Asymptotic Optimality of Massive MIMO Systems Using Densely Spaced Transmit Antennas}
%
%
%

\author{Keigo~Takeuchi,~\IEEEmembership{Member,~IEEE}
\thanks{
The material in this paper was submitted in part to 2016 IEEE 
International Symposium on Information Theory, Barcelona, Spain, Jul.\ 
2016~\cite{Takeuchi16}. 
}
\thanks{K.~Takeuchi is with the Department of Communication 
Engineering and Informatics, the University of Electro-Communications, 
Tokyo 182-8585, Japan (e-mail: ktakeuchi@uec.ac.jp).}
}

\markboth{IEEE transactions on information theory,~Vol.~, No.~,}%
{Takeuchi: Asymptotic Optimality of Massive MIMO Systems Using Densely Spaced Transmit Antennas}
%

\IEEEpubid{0000--0000/00\$00.00~\copyright~2016 IEEE}


\maketitle

\begin{abstract}
This paper considers a deterministic physical model of massive 
multiple-input multiple-output (MIMO) systems with uniform linear antenna 
arrays. It is known that the maximum spatial degrees of freedom is achieved 
by spacing antenna elements at half the carrier wavelength. 
The purpose of this paper is to investigate the impacts of spacing antennas  
more densely than the critical separation. 
The achievable rates of MIMO systems are evaluated in the large-system limit, 
where the lengths of transmit and receive antenna arrays tend to 
infinity with the antenna separations kept constant. 
The main results are twofold: One is that, under a mild assumption 
of channel instances, spacing antennas densely cannot improve the capacity 
of MIMO systems normalized by the spatial degrees of freedom. 
The other is that the normalized achievable rate of quadrature phase-shift 
keying converges to the normalized capacity achieved by optimal Gaussian 
signaling, as the transmit antenna separation tends to zero after taking 
the large-system limit. The latter result is based on mathematical 
similarity between MIMO transmission and faster-than-Nyquist signaling 
in signal space representations.  
\end{abstract}


%
\IEEEpeerreviewmaketitle

\section{Introduction}
\subsection{Motivation}
\IEEEPARstart{M}{assive} multiple-input multiple-output (MIMO) 
systems~\cite{Marzetta06,Marzetta10,Rusek13,Larsson14} are promising schemes  
for future wireless communications. In massive MIMO systems, very large  
antenna arrays are used to attain many spatial degrees of freedom. It is an 
important topic in information theory to elucidate the benefit obtained by 
utilizing such antenna arrays.  

Spatial correlations are a key factor that affects the performance of MIMO 
systems. In early work~\cite{Telatar99}, Telatar proved that, when the channel 
matrices have independent and identically distributed (i.i.d.) circularly 
symmetric complex Gaussian (CSCG) elements, the channel capacity of MIMO 
systems is proportional to the minimum of the numbers of transmit and receive 
antennas. However, experimental channel 
measurements~\cite{Wallace03,Chizhik03} demonstrated that this idealized 
assumption is broken in realistic MIMO systems, while the Gaussianity of 
each channel gain may be satisfied. Thus, the influence of spatial correlations 
has to be taken into account to understand the potential of MIMO systems. 

Spatial correlations are mainly caused by multipath fading and antenna 
properties, such as antenna spacing and mutual coupling between adjacent 
antenna elements~\cite{Jensen04}. Mutual coupling is a phenomenon 
that occurs between closely spaced antenna elements, and results in spatial 
correlations between transmit antennas and between receive 
antennas~\cite{Janaswamy02}. The influence of mutual coupling may be 
mitigated by constructing matching networks at both transmit and receive 
sides~\cite{Wallace04,Lau06,Volmer08}. Thus, this paper considers idealized 
MIMO systems with no mutual coupling, and focuses on the impacts of 
multipath fading and antenna spacing.    

For simplicity, consider uni-polarized uniform linear antenna arrays. 
Poon {\em et al.}~\cite{Poon05,Poon06} proved that the spatial degrees of 
freedom are at most $2\mathrm{min}\{L_{\mathrm{t}}, L_{\mathrm{r}}\}$ in general, 
in which $L_{\mathrm{t}}$ and $L_{\mathrm{r}}$ represent the lengths of the 
transmit and receive antenna arrays normalized by the carrier wavelength, 
respectively. When the normalized separations $\Delta_{\mathrm{t}}$ and 
$\Delta_{\mathrm{r}}$ of transmit and receive antenna elements are equal to 
the critical value $1/2$, 
the early result $\mathrm{min}\{L_{\mathrm{t}}/\Delta_{\mathrm{t}}, 
L_{\mathrm{r}}/\Delta_{\mathrm{r}}\}$ by Telatar~\cite{Telatar99} is consistent 
with the general result. Thus, the spatial degrees of freedom are dominated 
by the normalized array lengths.  

\IEEEpubidadjcol

In order to explain the motivation of this paper, we shall review the 
signal-space approach in \cite{Poon05}. A continuous transmit antenna array 
is analogous to a band-limited system with bandwidth $W=1$. 
The spatial domain $[-L_{\mathrm{t}}/2, L_{\mathrm{t}}/2]$ corresponds to the time 
domain, whereas the angular domain $[-1, 1]$ is associated with the 
frequency domain $[-W, W]$. The classical sampling theorem implies that, when 
$L_{\mathrm{t}}$ tends to infinity, any {\em continuous-time} signal in the 
spatial domain can be re-constructed by sampling the signal at antenna 
separation $\Delta_{\mathrm{t}}=1/(2W)$ corresponding to the Nyquist period. 
Thus, there are no points in spacing antenna elements more densely than the 
critically spaced case $\Delta_{\mathrm{t}}=1/2$, as long as continuous 
Gaussian signaling is used.    

The purpose of this paper is to investigate the influence of the transmit 
antenna separation for the case of suboptimal digital modulation. It has been 
shown that the performance can be improved by using a symbol period shorter 
than the critical Nyquist period, called faster-than-Nyquist (FTN) 
signaling~\cite{Mazo75,Rusek09,Yoo10}, when quadrature phase-shift keying 
(QPSK) is used. More precisely, Yoo and Cho~\cite{Yoo10} proved that 
QPSK FTN signaling\footnote{
Note that discrete FTN signaling does not contradict the sampling Theorem,  
which claims that there is a one-to-one correspondence between any 
continuous-time real-valued signal and the associated {\em real-valued} 
samples at the Nyquist rate. Discrete-valued samples do not necessarily 
represent all continuous-time real-valued signals. 

Instead of sending signals drawn from a memoryless Gaussian source, 
let us consider transmission of the corresponding quantized symbols on 
a finite alphabet. The rate-distortion theory~\cite{Cover06} implies that 
a diverging compression rate is required to achieve arbitrarily small 
mean-square-error distortion. This implies that a vanishing symbol 
period is needed in order for this scheme to achieve the transmission rate 
of Gaussian signaling for any signal-to-noise ratio.
} can achieve the channel capacity when the symbol period tends to zero. 
This result motivates us to investigate the densely spaced case 
$\Delta_{\mathrm{t}}<1/2$ for QPSK. 

\subsection{Methodology}
A well-established methodology for massive MIMO systems is the large-system 
analysis, in which the numbers of transmit and receive antennas are assumed to 
tend toward infinity at the same rate. In conventional large-system analysis, 
random matrix theory~\cite{Verdu99,Tse99,Tulino05,Dumont10,Couillet11,Mueller14} or the replica method~\cite{Tanaka02,Moustakas03,Guo05,Wen06,Zaidel12,Takeuchi12,Takeuchi13,Girnyk14} 
developed in statistical physics was utilized to analyze the performance of 
MIMO systems under the assumption of statistical channel models. 
The large-system analysis has been accepted from two points of view. 
One is that the convergence of the large-system limit is so quick that 
asymptotic results can provide good approximations for finite-sized 
systems~\cite{Biglieri02}. The other is that technological 
innovation~\cite{Rusek13,Larsson14} is increasing the number of antennas to 
be regarded as a realistic number, while there is still a limitation in the 
number of radio-frequency (RF) chains that can be equipped in massive MIMO 
transmitters. See e.g.\ \cite{Mesleh08,Jeganathan09,Renzo14,Takeuchi15} 
for a solution to the RF chain issue. 

We shall present a brief history of existing large-system analyses. 
It should be noted that it is of course impossible to review all efforts. 
In early works, channel matrices with i.i.d.\ 
zero-mean elements were investigated for Gaussian~\cite{Verdu99,Tse99} and 
non-Gaussian~\cite{Tanaka02,Guo05} signaling. Spatially correlated channel 
matrices were analyzed in \cite{Moustakas03,Tulino05} for Gaussian signaling. 
After the characterization of capacity-achieving input distributions 
for the spatially correlated case~\cite{Visotsky01,Jafar04,Tulino06}, 
the same methodology was applied to the precoder-design 
issue for Gaussian~\cite{Dumont10,Couillet11} and 
non-Gaussian~\cite{Wen06,Zaidel12,Girnyk14} signaling, as well as to the 
channel-estimation issue~\cite{Takeuchi12,Takeuchi13,Mueller14}. 

In order to investigate the influence of the transmit antenna separation, 
this paper considers a deterministic physical model of MIMO systems with 
uniform linear antenna arrays~\cite{Tse05}, rather than statistical channel 
models. All of the above existing works are based on statistical channel 
models. The conventional large-system limit corresponds to the limit in which 
the normalized lengths of the transmit and receive antenna arrays tend 
to infinity at the same rate with the normalized antenna separations 
kept constant. Thus, the latter limit---called large-system limit---is 
considered in this paper. 

The methodology presented in this paper is based on the notion of 
asymptotically equivalent matrices in the large-system limit. 
The notion provides an elementary proof of the classical Szeg\"o theorem 
for Toeplitz matrices~\cite{Gray06}. We compare the performance of massive 
MIMO systems with different antenna separations, by applying the same 
methodology to the corresponding channel matrices. 

\subsection{Contributions}
The main contributions of this paper are twofold. 
\begin{itemize}
\item We evaluate the achievable rate of massive MIMO systems with precoded 
Gaussian signaling---referred to as constrained capacity---normalized by the 
maximum spatial degrees of freedom. It is proved that there are no points in 
considering densely spaced antenna elements in terms of the normalized 
constrained capacity, with the exception of power gain. More precisely, 
the normalized constrained capacity for the densely spaced case is 
asymptotically equal to that for the critically spaced case 
under an appropriate scaling of signal-to-noise ratio (SNR). 

This statement itself is intuitively expected and proved in the high SNR 
regime~\cite{Poon06}. However, to the best of author's knowledge, 
no rigorous proof is known for finite SNRs. 

\item For non-Gaussian signaling, we consider the limit in which the 
transmit antenna separation tends to zero after taking the large-system 
limit. It is proved that the normalized achievable rate for precoded 
QPSK signaling converges to the corresponding constrained capacity 
achieved by precoded Gaussian signaling. This result is analogous to 
the asymptotic optimality of QPSK FTN signaling~\cite{Yoo10}. 

The result implies that the transmission 
scheme with bounded peak power can approximately achieve the constrained 
capacity for all SNRs, by using transmit antenna arrays with densely spaced 
antenna elements, whereas the peak power is unbounded for Gaussian signaling. 
Note that, due to precoding, the peak power tends to infinity in order to 
achieve the constrained capacity exactly. Thus, this statement does not 
contradict existing results with respect to peak power 
constraints~\cite{Smith71,Shamai95,Chan05}. 
\end{itemize}

\subsection{Organization}
The remainder of this paper is organized as follows: After presenting 
the notation used in this paper, in Section~\ref{sec2} the basic properties 
of uniform linear antenna arrays are reviewed. Technical key lemmas for 
proving the main results are also derived, while the proofs are given 
in Appendices~\ref{proof_lemma2} and \ref{proof_lemma3}. 

In Section~\ref{sec3} a deterministic physical model of MIMO systems is 
introduced. Furthermore, the angular domain representation of the channel 
model is reviewed. The representation is useful for proving the main results. 

Section~\ref{sec4} presents three main theorems. One of the main theorems 
is proved in Section~\ref{sec5}. The proofs of the other two theorems 
are given in Appendices~\ref{proof_theorem1} and \ref{proof_theorem2} on 
the basis of the notion of asymptotically equivalent matrices. After 
presenting numerical results in Section~\ref{sec6}, 
this paper is concluded in Section~\ref{sec7}. 

\subsection{Notation}
As basic parameters, we use the normalized length 
$L_{\mathrm{t}}$ of transmit antenna arrays, the normalized transmit antenna 
separation $\Delta_{\mathrm{t}}$, the number 
$M=L_{\mathrm{t}}/\Delta_{\mathrm{t}}$ of transmit antennas, the normalized 
length $L_{\mathrm{r}}$ of receive antenna arrays, the normalized receive  
antenna separation $\Delta_{\mathrm{r}}$, and the number 
$N=L_{\mathrm{r}}/\Delta_{\mathrm{r}}$ of receive antennas.   
The load $\alpha=L_{\mathrm{t}}/L_{\mathrm{r}}$ is defined as the ratio of 
the transmit array length to the receive array length, while 
{\em spurious} load $\beta=M/N$ is given by the ratio of the number of 
transmit antennas to the number of receive antennas. 

For integers $a, b\in\mathbb{Z}$ 
satisfying $a<b$, the set $\{a,a+1,\ldots,b-1\}$ of consecutive integers 
is simply represented as $[a: b)$, while $[a, b)$ denotes the interval for 
real $a, b\in\mathbb{R}$.  We define the sets $[a: b]$, $(a: b]$, and 
$(a: b)$ in the same manner. For any sequence $\{a_{n}\}$, the convention 
$\sum_{n\in\mathcal{N}}a_{n}=0$ is introduced when the set $\mathcal{N}$ of 
indices is empty. In this paper, the indices of vectors and matrices start  
from $0$. 

The imaginary unit is denoted by $j$. 
For a complex number $z\in\mathbb{C}$ and a complex matrix $\boldsymbol{A}$, 
the notations $z^{*}$, $\boldsymbol{A}^{\mathrm{T}}$, and 
$\boldsymbol{A}^{\mathrm{H}}$ represent the complex conjugate of $z$, 
the transpose of $\boldsymbol{A}$, and the conjugate transpose of 
$\boldsymbol{A}$, respectively. The matrix 
$\boldsymbol{E}_{n,k}(\boldsymbol{A})$ is defined as the extended matrix 
obtained by inserting $k$ all-zero columns and $k$ all-zero rows after the 
first $n$ columns and rows of $\boldsymbol{A}$, respectively, i.e.\ 
\begin{equation} \label{extended_matrix}
\boldsymbol{E}_{n,k}(\boldsymbol{A}) 
= \begin{pmatrix}
\boldsymbol{A}_{00} & \boldsymbol{O}_{n\times k} & \boldsymbol{A}_{01} \\ 
\boldsymbol{O}_{k\times n} & \boldsymbol{O}_{k\times k} & \boldsymbol{O} \\ 
\boldsymbol{A}_{10} & \boldsymbol{O} & \boldsymbol{A}_{11}  
\end{pmatrix}, 
\end{equation}
with 
\begin{equation}
\boldsymbol{A} 
= \begin{pmatrix}
\boldsymbol{A}_{00} & \boldsymbol{A}_{01} \\ 
\boldsymbol{A}_{10} & \boldsymbol{A}_{11} 
\end{pmatrix}, 
\quad \boldsymbol{A}_{00}\in\mathbb{C}^{n\times n}. 
\end{equation}

Let $\delta_{n,m}$ denote the Kronecker delta. 
The standard orthonormal basis of $\mathbb{R}^{N}$ is written as 
$\{\boldsymbol{e}_{N,n}\in\mathbb{R}^{N}| n\in[0: N)\}$, i.e.\ 
$(\boldsymbol{e}_{N,n})_{n'}=\delta_{n,n'}$. 
For a complex vector $\boldsymbol{v}$, 
the Euclidean norm is written as $\|\boldsymbol{v}\|$. 

The notation $\boldsymbol{a}\sim\mathcal{CN}(\boldsymbol{\mu},
\boldsymbol{\Sigma})$ indicates that the random vector $\boldsymbol{a}$ 
follows the proper complex Gaussian distribution with mean 
$\mathbb{E}[\boldsymbol{v}]=\boldsymbol{\mu}$ and 
covariance $\mathbb{E}[\boldsymbol{a}\boldsymbol{a}^{\mathrm{H}}]
=\boldsymbol{\Sigma}$. 
We use the standard notation~\cite{Cover06} for information-theoretical 
quantities such as mutual information.

\section{Uniform Linear Antenna Array} \label{sec2}
\subsection{Basic Properties} 
We review properties of uniform linear antenna 
arrays~\cite[Chapter~7]{Tse05} with no mutual coupling. 
Let $L$ and $\Delta$ denote the antenna length and the antenna separation 
normalized by the carrier wavelength, respectively. 
The number of antenna elements is given by $N=L/\Delta$, which must 
be a positive integer. Without loss of generality, we assume that $L$ is an 
integer. If $L$ is not, $L$ is replaced by $\lfloor L\rfloor$. 
This is equivalent to assuming a rational antenna 
separation $\Delta=L/N\in\mathbb{Q}$.  

Let $\boldsymbol{s}_{L,\Delta}(\Omega)\in\mathbb{C}^{N}$ denote the unit 
spatial signature with respect to directional cosine 
$\Omega=\cos\phi$ for $\phi\in[0, 2\pi)$  
\begin{equation} \label{signature}
\boldsymbol{s}_{L,\Delta}(\Omega) 
= \frac{1}{\sqrt{L/\Delta}}\left(
 1, e^{-2\pi j\Delta\Omega},\cdots, e^{-2\pi j(L/\Delta-1)\Delta\Omega}
\right)^{\mathrm{T}}.  
\end{equation}
The standard inner product between two vectors 
$\boldsymbol{s}_{L,\Delta}(\Omega)$ 
and $\boldsymbol{s}_{L,\Delta}(\Omega')$ depends on $\Omega$ and $\Omega'$ only 
through the difference $\Omega-\Omega'\in[-2, 2]$, 
and is given by 
\begin{IEEEeqnarray}{rl} 
f_{L,\Delta}(\Omega-\Omega') 
=& \boldsymbol{s}_{L,\Delta}(\Omega')^{\mathrm{H}}
\boldsymbol{s}_{L,\Delta}(\Omega) 
\nonumber \\ 
=& \frac{\Delta}{L}\sum_{n=0}^{L/\Delta-1}e^{-2\pi jn\Delta(\Omega-\Omega')}. 
\label{func_f}
\end{IEEEeqnarray}
Properties of the antenna array are characterized by 
the beamforming pattern $(\phi, |f_{L,\Delta}(\cos\phi_{0}-\cos\phi)|)$ 
with respect to direction $\phi_{0}\in[0, 2\pi)$.  

In order to investigate properties of the beamforming pattern, we extend 
the domain of $f_{L,\Delta}(\Omega)$ from $[-2, 2]$ to $\mathbb{R}$. 
The function $f_{L,\Delta}(\Omega)$ has the following 
properties~\cite{Tse05}: 
\begin{property} \label{property1}
\begin{itemize}
\item $f_{L,\Delta}(\Omega)$ is a periodic function with period $1/\Delta$. 
\item $f_{L,\Delta}(-\Omega)=f_{L,\Delta}^{*}(\Omega)$. Thus, 
$|f_{L,\Delta}(\Omega)|$ is even. 
\item For all integers $k\in\mathbb{Z}$, $f_{L,\Delta}(k/L)=1$ if $k$ is 
a multiple of $L/\Delta$. Otherwise, $f_{L,\Delta}(k/L)=0$. 
\item $|f_{L,\Delta}(\Omega)|\leq1$. 
\item  
\begin{equation} \label{func_f1} 
f_{L,\Delta}(\Omega) = e^{\pi j(L-\Delta)\Omega}
\mathrm{sinc}_{L/\Delta}(L\Omega),
\end{equation}
with
\begin{equation} \label{sinc}
\mathrm{sinc}_{N}(x) 
= \frac{1}{N}\frac{\sin(\pi x)}{\sin(\pi x/N)},  
\end{equation}
where we define $\mathrm{sinc}_{N}(x)=1$ at $x=kN$ for all $k\in\mathbb{Z}$. 
\end{itemize}
\end{property} 

Note that the behavior of (\ref{sinc}) is similar to that of the sinc 
function $\mathrm{sinc}(x)=\sin(\pi x)/\pi x$. In fact, we have the 
point-wise convergence  
$\lim_{N\to\infty}\mathrm{sinc}_{N}(x)=\mathrm{sinc}(x)$. Furthermore, 
the following upper bound holds:  
\begin{equation} \label{sinc_bound}
|\mathrm{sinc}_{N}(x)|\leq\frac{1}{2x}, 
\quad \hbox{for all $x\in(0,N/2]$}, 
\end{equation}
since 
\begin{equation}
x|\mathrm{sinc}_{N}(x)| 
\leq \frac{1}{\pi}\frac{\pi x/N}{\sin(\pi x/N)}
\leq \frac{1}{2},
\end{equation}
where the last inequality is due to the fact that $u/\sin(u)$ is monotonically 
increasing for $u\in(0, \pi/2]$. 

\subsection{Basis Expansion}
We next introduce a basis expansion of the spatial signature. 
The third property in Property~\ref{property1} implies that 
the signature vectors 
$\{\boldsymbol{s}_{L,\Delta}(k/L)| k\in[0: L/\Delta)\}$ form an 
orthonormal basis of $\mathbb{C}^{N}$. Representing the spatial signature 
$\boldsymbol{s}_{L,\Delta}(\Omega)$ with the basis yields 
\begin{equation} \label{basis_expansion}
\boldsymbol{s}_{L,\Delta}(\Omega) 
= \sum_{k=0}^{L/\Delta-1}f_{L,\Delta}\left(
 \frac{k}{L} - \Omega
\right) 
\boldsymbol{s}_{L,\Delta}\left(
 \frac{k}{L} 
\right). 
\end{equation} 

\begin{figure}[t]
\begin{center}
\includegraphics[width=\hsize]{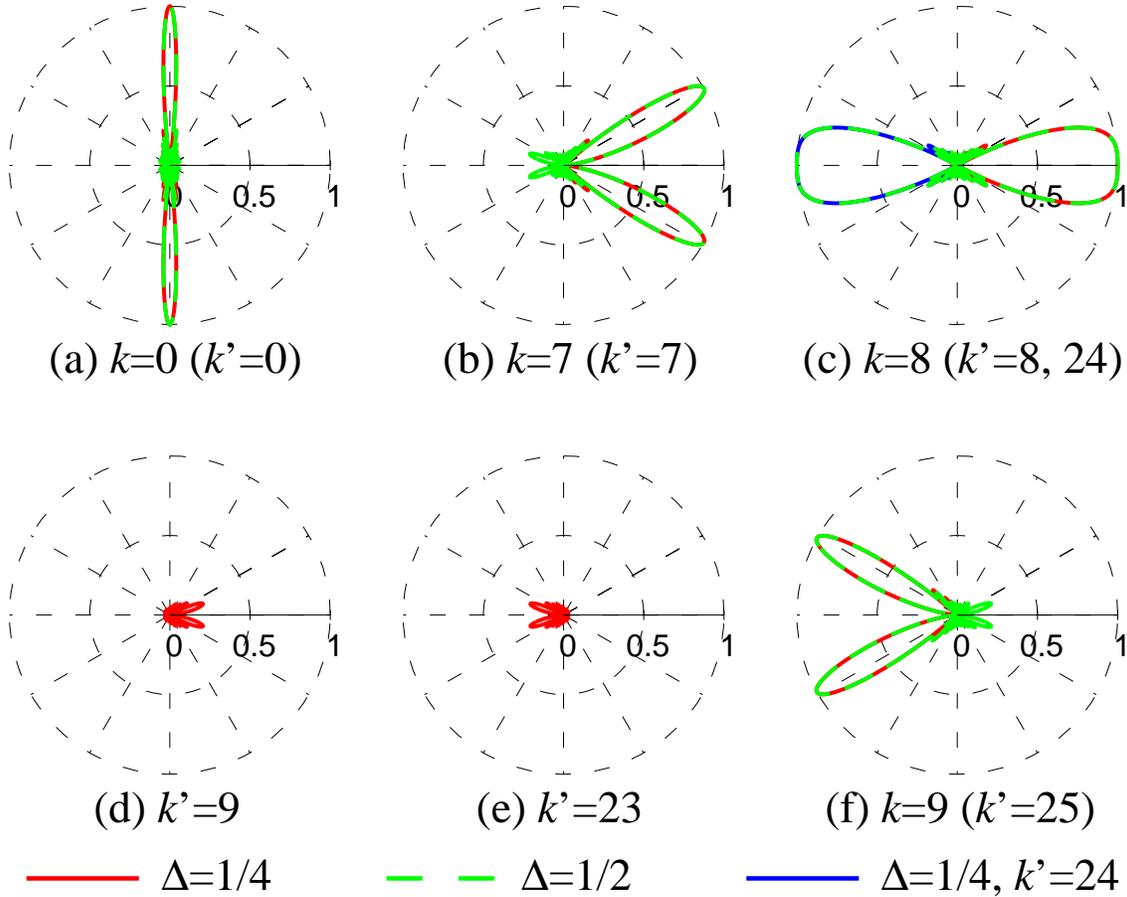}
\caption{
Beamforming patterns $(\phi, |f_{L,\Delta}(k/L-\cos\phi)|)$ for the critically 
spaced case $\Delta=1/2$, and $(\phi, |f_{L,\Delta}(k'/L-\cos\phi)|)$ 
for the densely spaced case $\Delta=1/4$. The normalized length of the 
antenna array is set to $L=8$.   
}
\label{fig1}
\end{center}
\end{figure}

Figure~\ref{fig1} shows the properties of $f_{L,\Delta}(k/L-\Omega)$ for 
the critically spaced case $\Delta=1/2$ and the densely spaced case 
$\Delta<1/2$. There is approximate correspondence between 
$f_{L,1/2}(k/L-\Omega)$ and $f_{L,\Delta}(k/L-\Omega)$ with $\Delta<1/2$ for 
$k\in[0: L)$, and between $f_{L,1/2}(k/L-\Omega)$ and 
$f_{L,\Delta}((k+L/\Delta-2L)/L-\Omega)$ for $k\in(L: 2L)$. Furthermore, 
the beamforming pattern for $k=L$ and $\Delta=1/2$ approximately coincides 
with a superposition of those for $k=L$ and $k=L/\Delta-L$ in the densely 
spaced case $\Delta<1/2$. The beamforming patterns for the densely spaced 
case have no main lobes when $k\in(L: L/\Delta -L)$.  

We shall present three technical key lemmas to justify these observations. 
A first lemma implies that channel gains introduced in the next section 
are adequately normalized. 

\begin{lemma} \label{lemma1}
For $\Omega, \Omega'\in[-1,1]$, define 
\begin{equation} \label{sum}
I_{1}(\Omega,\Omega'; L,\Delta) 
= \sum_{k=0}^{L/\Delta-1}
f_{L,\Delta}\left(
 \frac{k}{L} - \Omega
\right)
f_{L,\Delta}^{*}\left(
 \frac{k}{L} - \Omega'
\right).  
\end{equation}
Then, 
\begin{equation} \label{sum_ans}
I_{1}(\Omega,\Omega';L,\Delta) 
= f_{L,\Delta}(\Omega'-\Omega). 
\end{equation}
In particular, $|I_{1}(\Omega,\Omega';L,\Delta)|\leq1$ holds. 
\end{lemma}
\begin{IEEEproof}
Note $N=L/\Delta$. Substituting (\ref{func_f}) into (\ref{sum}) yields 
\begin{IEEEeqnarray}{rl}
I_{1}
=& \frac{1}{N^{2}}\sum_{n=0}^{N-1}\sum_{n'=0}^{N-1}
e^{2\pi j\Delta(n\Omega - n'\Omega')}
\sum_{k=0}^{N-1}e^{-2\pi j(n-n')k/N} 
\nonumber \\ 
=& \frac{1}{N}\sum_{n=0}^{N-1}e^{2\pi jn\Delta(\Omega - \Omega')}, 
\end{IEEEeqnarray}
which implies (\ref{sum_ans}) from (\ref{func_f}).  
The last statement in the lemma holds from the fourth property in 
Property~\ref{property1}. 
\end{IEEEproof}

In order to present the remaining two lemmas, we decompose the set 
$[0:L/\Delta)$ of antenna indices into the disjoint subsets 
$\{L\}\cup\mathcal{K}_{1}\cup\mathcal{K}_{2}$, with 
\begin{IEEEeqnarray}{l}
\mathcal{K}_{1}=[0:L)\cup[L/\Delta-L:L/\Delta), 
\nonumber \\ 
\mathcal{K}_{2}=(L:L/\Delta-L).
\end{IEEEeqnarray} 
For $\Delta=1/2$, we define $\mathcal{K}_{1}=[0:L/\Delta)$ and 
$\mathcal{K}_{2}=\emptyset$. 

For $l\in\mathbb{N}$, let $\{s_{l}\in\mathbb{N}\}$ denote a sequence of 
positive integers that satisfies two conditions: 
$\lim_{l\to\infty}s_{l}=\infty$ and $\lim_{l\to\infty}s_{l}/l=0$. 
We restrict the domain of the directional cosine $\Omega$ from $[-1, 1]$ 
to $\mathcal{D}_{L}=[-(1-s_{L}/L), 1-s_{L}/L]$. The neighborhoods of the 
boundaries in the angular domain $[-1 ,1]$ contribute to neighborhoods of the 
two indices $L$ and $L/\Delta-L$ in the spatial domain. Since the contribution 
is small but non-negligible, we consider the restricted interval 
$\mathcal{D}_{L}$ in the angular domain.  

\begin{lemma} \label{lemma2}
For $\Omega,\Omega'\in\mathcal{D}_{L}$, define
\begin{equation} \label{partial_sum} 
I_{2}(\Omega,\Omega';L,\Delta) 
= \sum_{k\in\mathcal{K}_{2}}
f_{L,\Delta}\left(
 \frac{k}{L} - \Omega
\right)
f_{L,\Delta}^{*}\left(
 \frac{k}{L} - \Omega'
\right).  
\end{equation}
Then, there exists some constant $A>0$ such that 
\begin{equation}
s_{L}|I_{2}(\Omega,\Omega';L,\Delta)| < A, 
\end{equation} 
for all $\Omega, \Omega'\in\mathcal{D}_{L}$ and all $\Delta\in(0,1/2]$ 
in the limit $L, N\to\infty$ with $\Delta=L/N$ fixed. 
\end{lemma}
\begin{IEEEproof}
See Appendix~\ref{proof_lemma2}. 
\end{IEEEproof}

\begin{lemma} \label{lemma3}
Let the one-to-one mapping $\kappa_{L,\Delta}(k)$ from $[0:2L]$ onto 
$\mathcal{K}_{1}\cup\{L\}$ as 
\begin{equation} \label{kappa}
\kappa_{L,\Delta}(k) 
= \left\{
\begin{array}{cl}
k & \hbox{for $k\in[0:L)$,} \\ 
k + L/\Delta - 2L& \hbox{for $k\in[L:2L)$,} \\
L & \hbox{for $k=2L$.}
\end{array}
\right.
\end{equation}
For $\Omega, \Omega'\in\mathcal{D}_{L}$, define 
\begin{IEEEeqnarray}{rl} 
I_{3}(\Omega,\Omega';L,\Delta) 
=& \sum_{k=0}^{2L-1}D_{k,\kappa_{L,\Delta}(k)}(\Omega)
D_{k,\kappa_{L,\Delta}(k)}^{*}(\Omega') 
\nonumber \\ 
&+ f_{L,\Delta}(1-\Omega)f_{L,\Delta}^{*}(1-\Omega'),
\label{difference_sum} 
\end{IEEEeqnarray}
with 
\begin{equation} \label{D_k0}
D_{k,k'}(\Omega)
= f_{L,1/2}\left(
 \frac{k}{L} - \Omega
\right)
- f_{L,\Delta}\left(
 \frac{k'}{L} - \Omega
\right). 
\end{equation}
Then, there exists some constant $A>0$ such that 
\begin{equation}
s_{L}|I_{3}(\Omega,\Omega';L,\Delta)|<A, 
\end{equation}
for all $\Omega, \Omega'\in\mathcal{D}_{L}$ and all $\Delta\in(0,1/2]$ 
in the limit $L, N\to\infty$ with $\Delta=L/N$ fixed.   
\end{lemma}
\begin{IEEEproof}
See Appendix~\ref{proof_lemma3}. 
\end{IEEEproof}

All lemmas depend highly on the properties of the function 
$f_{L,\Delta}(\Omega)$. Lemma~\ref{lemma2} 
indicates that most of the energy is concentrated on the antenna 
indices $\mathcal{K}_{1}\cup\{L\}$. Lemma~\ref{lemma3} will be used to 
evaluate the capacity difference between the critically spaced case and 
the densely spaced case. 

\section{Channel Model} \label{sec3}
\subsection{MIMO Channel} 
Consider MIMO channels with $M$ transmit antennas and $N$ receive antennas. 
The received vector $\boldsymbol{y}\in\mathbb{C}^{N}$ is given by 
\begin{equation} \label{MIMO} 
\boldsymbol{y} 
= \sqrt{\gamma}\boldsymbol{H}\boldsymbol{x} + \boldsymbol{w}. 
\end{equation}
In (\ref{MIMO}), $\boldsymbol{H}\in\mathbb{C}^{N\times M}$, 
$\boldsymbol{x}\in\mathbb{C}^{M}$, and 
$\boldsymbol{w}\sim\mathcal{CN}(\boldsymbol{0},\boldsymbol{I}_{N})$ 
denote the channel matrix, the transmitted vector, and the 
additive white Gaussian noise (AWGN) vector, 
respectively. The average power constraint 
$\mathbb{E}[\|\boldsymbol{x}\|^{2}]\leq 1$ is imposed. 
The parameter $\gamma>0$ corresponds to the SNR. 

\subsection{Physical Modeling} 
Uniform linear antenna arrays with no mutual coupling are assumed. 
Let $\Delta_{\mathrm{t}}$ and $\Delta_{\mathrm{r}}$ denote the transmit and 
receive antenna separation normalized by the carrier wavelength, respectively. 
The normalized lengths of the transmit and receive antenna arrays are given 
by $L_{\mathrm{t}}=M\Delta_{\mathrm{t}}$ and $L_{\mathrm{r}}=N\Delta_{\mathrm{r}}$. 
A deterministic physical model~\cite[Eq.~(7.56)]{Tse05} of the channel matrix 
$\boldsymbol{H}\in\mathbb{C}^{N\times M}$ is given by 
\begin{equation} \label{channel} 
\boldsymbol{H} 
= \sqrt{NM}\int a(p)\boldsymbol{s}_{L_{\mathrm{r}},\Delta_{\mathrm{r}}}
(\Omega_{\mathrm{r}}(p))
\boldsymbol{s}_{L_{\mathrm{t}},\Delta_{\mathrm{t}}}
(\Omega_{\mathrm{t}}(p))^{\mathrm{H}}dp,  
\end{equation} 
where the transmit and receive unit spatial signature vectors  
$\boldsymbol{s}_{L_{\mathrm{t}},\Delta_{\mathrm{t}}}(\Omega)\in\mathbb{C}^{M}$ and 
$\boldsymbol{s}_{L_{\mathrm{r}},\Delta_{\mathrm{r}}}(\Omega)\in\mathbb{C}^{N}$ with 
respect to directional cosine $\Omega$ are defined by (\ref{signature}). 
In (\ref{channel}), $a(p)\in\mathbb{C}$ represents the complex attenuation 
of path~$p$. The directional cosines  
$\Omega_{\mathrm{t}}(p)=\cos\phi_{\mathrm{t}}(p)\in[-1, 1]$ and 
$\Omega_{\mathrm{r}}(p)=\cos\phi_{\mathrm{r}}(p)\in[-1, 1]$ are defined 
via the departure angle $\phi_{\mathrm{t}}(p)$ from the transmit antenna array 
and via the incidence angle $\phi_{\mathrm{r}}(p)$ to the receive antenna 
array for path~$p$. 

\subsection{Angular Domain Representation}
We next introduce the angular domain representation of the channel 
matrix $\boldsymbol{H}$. 
Substituting the basis expansions~(\ref{basis_expansion}) for 
$\boldsymbol{s}_{L_{\mathrm{t}},\Delta_{\mathrm{t}}}(\Omega_{\mathrm{t}}(p))$ and 
$\boldsymbol{s}_{L_{\mathrm{r}},\Delta_{\mathrm{r}}}(\Omega_{\mathrm{r}}(p))$ 
into (\ref{channel}), 
we find that the channel matrix~(\ref{channel}) can be represented as 
\begin{equation} \label{tilde_channel} 
\sqrt{\gamma}\boldsymbol{H} 
= \sqrt{\tilde{\gamma}}\boldsymbol{U}_{L_{\mathrm{r}},\Delta_{\mathrm{r}}}
\boldsymbol{G}_{\Delta_{\mathrm{t}},\Delta_{\mathrm{r}}}
\boldsymbol{U}_{L_{\mathrm{t}},\Delta_{\mathrm{t}}}^{\mathrm{H}}, 
\end{equation}
with 
\begin{equation} \label{normalized_SNR}
\tilde{\gamma}
= \frac{\gamma}{4\Delta_{\mathrm{t}}\Delta_{\mathrm{r}}}.
\end{equation}
In (\ref{tilde_channel}), the $N\times N$ unitary matrix 
$\boldsymbol{U}_{L_{\mathrm{r}},\Delta_{\mathrm{r}}}$ has 
$\boldsymbol{s}_{L_{\mathrm{r}},\Delta_{\mathrm{r}}}(n/L_{\mathrm{r}})$ as the $n$th column 
for $n=0,\ldots,N-1$, while the $M\times M$ unitary matrix 
$\boldsymbol{U}_{L_{\mathrm{t}},\Delta_{\mathrm{t}}}$ has 
$\boldsymbol{s}_{L_{\mathrm{t}},\Delta_{\mathrm{t}}}(m/L_{\mathrm{t}})$ 
as the $m$th column for $m=0,\ldots,M-1$. 
Note that the $\boldsymbol{U}_{L_{\mathrm{t}},\Delta_{\mathrm{t}}}$ and 
$\boldsymbol{U}_{L_{\mathrm{r}},\Delta_{\mathrm{r}}}$ are equal to the 
$M$-point and $N$-point discrete Fourier transform (DFT) matrices, 
respectively. 

The $(n, m)$ entry $g_{n,m}$ of the 
channel matrix $\boldsymbol{G}_{\Delta_{\mathrm{t}},\Delta_{\mathrm{r}}}
\in\mathbb{C}^{N\times M}$ in the angular domain is given by 
\begin{IEEEeqnarray}{rl} 
g_{n,m}
= \sqrt{4L_{\mathrm{t}}L_{\mathrm{r}}}
\int a(p)&f_{L_{\mathrm{r}},\Delta_{\mathrm{r}}}\left(
 \frac{n}{L_{\mathrm{r}}} - \Omega_{\mathrm{r}}(p)
\right)
\nonumber \\ 
\cdot&f_{L_{\mathrm{t}},\Delta_{\mathrm{t}}}^{*}\left(
 \frac{m}{L_{\mathrm{t}}} - \Omega_{\mathrm{t}}(p)
\right)dp. \label{channel_gain}
\end{IEEEeqnarray}

The prefactor $1/(4\Delta_{\mathrm{t}}\Delta_{\mathrm{r}})\geq1$ in the 
normalized SNR~$\tilde{\gamma}$ represents the power gain obtained by spacing 
antenna elements densely. Since we have ignored a power loss due to mutual 
coupling, we will exclude the influence of the power gain 
$1/(4\Delta_{\mathrm{t}}\Delta_{\mathrm{r}})$ in comparing the critically 
spaced case and the densely spaced case. In other words, the normalized 
SNR $\tilde{\gamma}$ will be fixed in comparisons between the two cases. 
This implies that the actual SNRs $\gamma$ are different from each other 
in the two cases.  

We consider the large-system limit, in which $N$, $M$, $L_{\mathrm{t}}$, and 
$L_{\mathrm{r}}$ tend to infinity with the ratios 
$\Delta_{\mathrm{t}}=L_{\mathrm{t}}/M$, $\Delta_{\mathrm{r}}=L_{\mathrm{r}}/N$, 
and $\alpha=L_{\mathrm{t}}/L_{\mathrm{r}}$ kept constant.  
Throughout this paper, we postulate the following for the deterministic 
channel instance $\mathcal{C}=\{a(\cdot), \Omega_{\mathrm{r}}(\cdot), 
\Omega_{\mathrm{t}}(\cdot)\}$.  

\begin{assumption} \label{assumption1}
Let $\mathcal{D}_{l}=[-(1-s_{l}/l), 1-s_{l}/l]$, in which 
$\{s_{l}\in\mathbb{N}\}$ denotes a slowly diverging sequence of positive 
integers that satisfies $\lim_{l\to\infty}s_{l}=\infty$ and 
$\lim_{l\to\infty}s_{l}/l=0$. Postulate a class $\mathfrak{C}$ of 
channel instances (a set of $\mathcal{C}$) satisfying that 
$\Omega_{\mathrm{r}}(\cdot)\in\mathcal{D}_{L_{\mathrm{r}}}$ and 
$\Omega_{\mathrm{t}}(\cdot)\in\mathcal{D}_{L_{\mathrm{t}}}$ hold, and that the total 
power $\int|a(p)|^{2}dp$ of attenuation and the maximum singular value of 
$\mathrm{min}\{2L_{\mathrm{t}}, 2L_{\mathrm{r}}\}^{-1/2}
\boldsymbol{G}_{\Delta_{\mathrm{t}},\Delta_{\mathrm{r}}}$ are 
uniformly bounded in the large-system limit for all channel instances 
$\mathcal{C}\in\mathfrak{C}$. 
\end{assumption}

The angular domain is restricted in order to use Lemmas~\ref{lemma2} and 
\ref{lemma3}. If the channel instances are sampled from proper statistical 
models, $\Omega_{\mathrm{r}}(\cdot)$ and $\Omega_{\mathrm{t}}(\cdot)$ should be 
almost surely included into the restricted intervals in the large-system limit. 


The bounded maximum singular value implies that we can enjoy no 
{\em noiseless} eigen channels in the large-system limit for finite SNR. 
Thus, this assumption should be satisfied for practical MIMO channels. 

\section{Main Results} \label{sec4}
\subsection{Constrained Capacity} 
Let $\boldsymbol{Q}_{M}$ denote an $M\times M$ covariance matrix 
satisfying the power constraint $\mathrm{Tr}(\boldsymbol{Q}_{M})\leq1$. 
We consider precoded Gaussian signaling 
$\boldsymbol{x}\sim\mathcal{CN}(\boldsymbol{0},\boldsymbol{Q}_{M})$. 
It is well known that the precoding scheme achieves the constrained capacity 
of the MIMO channel~(\ref{MIMO}) with the channel 
matrix $\boldsymbol{H}$ given by (\ref{tilde_channel}),   
\begin{IEEEeqnarray}{rl} 
&C_{\mathrm{opt}}(\boldsymbol{Q}_{M};\tilde{\gamma}, 
\boldsymbol{G}_{\Delta_{\mathrm{t}},\Delta_{\mathrm{r}}})  
\nonumber \\ 
=& \log\det\left(
 \boldsymbol{I}  
 + \tilde{\gamma}\boldsymbol{G}_{\Delta_{\mathrm{t}},\Delta_{\mathrm{r}}}
 \boldsymbol{U}_{L_{\mathrm{t}},\Delta_{\mathrm{t}}}^{\mathrm{H}}
 \boldsymbol{Q}_{M}\boldsymbol{U}_{L_{\mathrm{t}},\Delta_{\mathrm{t}}}
 \boldsymbol{G}_{\Delta_{\mathrm{t}},\Delta_{\mathrm{r}}}^{\mathrm{H}}
\right). \label{capacity}
\end{IEEEeqnarray}
The channel capacity is equal to the maximum of 
(\ref{capacity}) over all possible covariance matrices 
$\boldsymbol{Q}_{M}$ satisfying the average power constraint 
$\mathrm{Tr}(\boldsymbol{Q}_{M})\leq 1$. 

The degree of spatial freedom is at most 
$2\mathrm{min}\{L_{\mathrm{t}},L_{\mathrm{r}}\}$~\cite{Poon05}. 
When $\Delta_{\mathrm{t}}, \Delta_{\mathrm{r}}<1/2$, 
most of the power of the channel gains $g_{n,m}$ should be concentrated on 
$n\in\mathcal{N}=[0:L_{\mathrm{r}}]\cup[N-L_{\mathrm{r}}:N)$ and 
$m\in\mathcal{M}=[0:L_{\mathrm{t}}]\cup[M-L_{\mathrm{t}}:M)$ 
in the large-system limit. For the critically spaced 
case $\Delta_{\mathrm{t}}=\Delta_{\mathrm{r}}=1/2$, all channel gains should 
have significant power. 

In order to present a precise statement for this intuition, we consider the 
densely spaced case 
$\Delta_{\mathrm{t}}, \Delta_{\mathrm{r}}<1/2$, and define the $N\times M$ 
matrix $\tilde{\boldsymbol{G}}_{\Delta_{\mathrm{t}}, \Delta_{\mathrm{r}}}$ as 
\begin{equation} \label{hat_G}
(\tilde{\boldsymbol{G}}_{\Delta_{\mathrm{t}}, \Delta_{\mathrm{r}}})_{n,m} 
= \left\{
\begin{array}{cl}
g_{n,m} & \hbox{for $(n, m)\in\mathcal{N}\times\mathcal{M}$,} \\ 
0 & \mathrm{otherwise.}
\end{array}
\right.
\end{equation}
Furthermore, it is convenient to define the covariance matrix 
$\boldsymbol{\Sigma}_{M}
=\boldsymbol{U}_{L_{\mathrm{t}},\Delta_{\mathrm{t}}}^{\mathrm{H}}
\boldsymbol{Q}_{M}\boldsymbol{U}_{L_{\mathrm{t}},\Delta_{\mathrm{t}}}$. 

\begin{assumption} \label{assumption2}
Postulate the set $\mathfrak{S}_{M}$ of $M\times M$ covariance matrices 
$\boldsymbol{\Sigma}_{M}$ in which the power constraint 
$\mathrm{Tr}(\boldsymbol{\Sigma}_{M})\leq1$ is satisfied, and in which  
the maximum eigenvalue of $2L_{\mathrm{t}}\boldsymbol{\Sigma}_{M}$ is 
uniformly bounded in the large-system limit. 
\end{assumption}

\begin{theorem} \label{theorem1} 
Suppose that the precoding matrix is given by  
$\boldsymbol{Q}_{M}=\boldsymbol{U}_{L_{\mathrm{t}},\Delta_{\mathrm{t}}}
\boldsymbol{\Sigma}_{M}\boldsymbol{U}_{L_{\mathrm{t}},\Delta_{\mathrm{t}}}^{\mathrm{H}}$. 
Fix SNR $\tilde{\gamma}>0$, receive antenna separation 
$\Delta_{\mathrm{r}}\in(0,1/2]$, and load $\alpha=L_{\mathrm{t}}/L_{\mathrm{r}}>0$. 
Under Assumptions~\ref{assumption1} and \ref{assumption2}, 
the following limit 
\begin{equation} \label{asym_capacity}
\frac{|C_{\mathrm{opt}}(\boldsymbol{Q}_{M};\tilde{\gamma}, 
 \boldsymbol{G}_{\Delta_{\mathrm{t}}, \Delta_{\mathrm{r}}})
 - C_{\mathrm{opt}}(\boldsymbol{Q}_{M};\tilde{\gamma}, 
 \tilde{\boldsymbol{G}}_{\Delta_{\mathrm{t}}, \Delta_{\mathrm{r}}})|
}{2\min\{L_{\mathrm{t}},L_{\mathrm{r}}\}} \to0
\end{equation}
holds uniformly for all transmit antenna separations 
$\Delta_{\mathrm{t}}\in(0,1/2]$,  
covariance matrices $\boldsymbol{\Sigma}_{M}\in\mathfrak{S}_{M}$, 
and all channel instances $\mathcal{C}\in\mathfrak{C}$ in the large-system 
limit. 
\end{theorem}
\begin{IEEEproof}
See Appendix~\ref{proof_theorem1}. 
\end{IEEEproof}

Theorem~\ref{theorem1} implies that most of the power of the channel gains 
is concentrated on the antenna indices $\mathcal{N}\times\mathcal{M}$ 
in terms of the constrained capacity normalized by the degree of spatial 
freedom. Thus, it is sufficient to consider power allocation over 
$m\in\mathcal{M}$, as long as the large-system limit is taken. 
In other words, we consider the $M\times M$ covariance matrix 
$\boldsymbol{E}_{L_{\mathrm{t}}+1,M-(2L_{\mathrm{t}}+1)}
(\boldsymbol{\Sigma}_{2L_{\mathrm{t}}+1})$ defined via (\ref{extended_matrix}) 
for $\boldsymbol{\Sigma}_{2L_{\mathrm{t}}+1}\in\mathfrak{S}_{2L_{\mathrm{t}}+1}$. 

We next make a comparison between the critically spaced case and the 
densely spaced case with identical $L_{\mathrm{t}}$ and $L_{\mathrm{r}}$, 
so that the numbers of antennas are different from each other for the 
two cases.

\begin{theorem} \label{theorem2}
Let $\boldsymbol{Q}_{2L_{\mathrm{t}}}=\boldsymbol{U}_{L_{\mathrm{t}},1/2}
\boldsymbol{\Sigma}_{2L_{\mathrm{t}}}\boldsymbol{U}_{L_{\mathrm{t}},1/2}^{\mathrm{H}}$ 
and $\boldsymbol{Q}_{M}=\boldsymbol{U}_{L_{\mathrm{t}},\Delta_{\mathrm{t}}}
\boldsymbol{E}_{L_{\mathrm{t}}+1,M-(2L_{\mathrm{t}}+1)}(\boldsymbol{\Sigma}_{2L_{\mathrm{t}}+1})
\boldsymbol{U}_{L_{\mathrm{t}},\Delta_{\mathrm{t}}}^{\mathrm{H}}$, defined via 
(\ref{extended_matrix}).  
Assume that 
\begin{equation} \label{condition}
\mathrm{Tr}\left\{
 \left(
  \boldsymbol{E}_{L_{\mathrm{t}},1}(\boldsymbol{\Sigma}_{2L_{\mathrm{t}}})
  - \boldsymbol{\Sigma}_{2L_{\mathrm{t}}+1}
 \right)^{2}
\right\}\to0 
\end{equation}
as $L_{\mathrm{t}}\to\infty$. 
Under Assumptions~\ref{assumption1} and \ref{assumption2},  
for fixed SNR $\tilde{\gamma}>0$ and load $\alpha>0$ 
the following limit  
\begin{equation} \label{capacity_difference}
\frac{
 |C_{\mathrm{opt}}(\boldsymbol{Q}_{2L_{\mathrm{t}}};\tilde{\gamma},
  \boldsymbol{G}_{1/2,1/2})
 - C_{\mathrm{opt}}(\boldsymbol{Q}_{M};\tilde{\gamma}, 
 \tilde{\boldsymbol{G}}_{\Delta_{\mathrm{t}}, \Delta_{\mathrm{r}}})|
}{2\min\{L_{\mathrm{t}},L_{\mathrm{r}}\}}\to0 
\end{equation}
holds uniformly for all antenna separations 
$\Delta_{\mathrm{t}}, \Delta_{\mathrm{r}}\leq1/2$, 
covariance matrices $\boldsymbol{\Sigma}_{2L_{\mathrm{t}}}
\in\mathfrak{S}_{2L_{\mathrm{t}}}$, 
$\boldsymbol{\Sigma}_{2L_{\mathrm{t}}+1}\in\mathfrak{S}_{2L_{\mathrm{t}}+1}$, 
and all channel instances $\mathcal{C}\in\mathfrak{C}$ 
in the large-system limit. 
\end{theorem}
\begin{IEEEproof}
See Appendix~\ref{proof_theorem2}. 
\end{IEEEproof}

From Theorems~\ref{theorem1} and \ref{theorem2}, we conclude that the 
normalized capacity for the critically spaced case is equal to that 
for the densely spaced case 
in the large-system limit. In other words, there are no points in using 
uniform linear antenna arrays with densely spaced antenna elements for 
all SNRs, as long as optimal Gaussian signaling is assumed. To the best of 
author's knowledge, Theorems~\ref{theorem1} and \ref{theorem2} are the first 
theoretical results for finite SNRs, although the optimality of the critically 
spaced case was proved in the high SNR limit~\cite{Poon06}.  

\subsection{Non-Gaussian Signaling}
We have so far shown the asymptotic optimality of the critically spaced case 
for Gaussian signaling. The purpose of this section is to investigate what 
occurs for suboptimal non-Gaussian signaling, such as QPSK. 

Let $\boldsymbol{b}=(b_{0},\ldots,b_{M-1})^{\mathrm{T}}$ denote the 
$M$-dimensional data symbol vector that has independent QPSK elements 
$\{b_{m}\}$ with unit power. 
For a square root $\boldsymbol{Q}_{M}^{1/2}$ of an $M\times M$ covariance 
matrix $\boldsymbol{Q}_{M}\in\mathfrak{S}_{M}$, the transmitted vector 
$\boldsymbol{x}=\boldsymbol{Q}_{M}^{1/2}\boldsymbol{b}$ is 
generated as the product of the precoding matrix 
$\boldsymbol{Q}_{M}^{1/2}$ and the symbol vector $\boldsymbol{b}$. 
Substituting $\boldsymbol{x}=\boldsymbol{Q}_{M}^{1/2}\boldsymbol{b}$ and 
(\ref{tilde_channel}) into (\ref{MIMO}) yields 
\begin{equation} \label{effective_channel}
\boldsymbol{y} 
= \sqrt{\tilde{\gamma}}\boldsymbol{A}\boldsymbol{b} + \boldsymbol{w}, 
\end{equation}
where the effective channel matrix 
$\boldsymbol{A}\in\mathbb{C}^{N\times M}$ is given by  
\begin{equation} \label{effective_channel_matrix}
\boldsymbol{A} 
= \boldsymbol{U}_{L_{\mathrm{r}},\Delta_{\mathrm{r}}}
\boldsymbol{G}_{\Delta_{\mathrm{t}},\Delta_{\mathrm{r}}}
\boldsymbol{U}_{L_{\mathrm{t}},\Delta_{\mathrm{t}}}^{\mathrm{H}}
\boldsymbol{Q}_{M}^{1/2}. 
\end{equation}
Then, the achievable rate 
$C(\boldsymbol{Q}_{M};\tilde{\gamma},
\boldsymbol{G}_{\Delta_{\mathrm{t}},\Delta_{\mathrm{r}}})$ 
of the precoded QPSK scheme is defined as 
\begin{equation} \label{QPSK_capacity} 
C(\boldsymbol{Q}_{M};\tilde{\gamma},
\boldsymbol{G}_{\Delta_{\mathrm{t}},\Delta_{\mathrm{r}}}) 
= I(\boldsymbol{b};\boldsymbol{y}), 
\end{equation}  
where the deterministic channel matrix $\boldsymbol{A}$ is fixed. 

The main result for QPSK is that the achievable 
rate~(\ref{QPSK_capacity}) normalized by the degree of spatial freedom 
converges to the normalized constrained capacity in the dense limit 
$\Delta_{\mathrm{t}}\to0$ after taking the large-system limit. 

\begin{assumption} \label{assumption3}
Postulate the set $\tilde{\mathfrak{S}}_{M}$ of $M\times M$ covariance 
matrices $\boldsymbol{Q}_{M}$, in which $\tilde{\mathfrak{S}}_{M}$ is a 
subset of $\mathfrak{S}_{M}$ defined in Assumption~\ref{assumption2}, 
and in which there exists some constant $A>0$ such that  
$\|(2L_{\mathrm{t}})^{1/2}\boldsymbol{Q}_{M}^{1/2}\boldsymbol{e}_{M,m}\|^{2}
\leq A\Delta_{\mathrm{t}}$ holds for all $m\in[0:M)$.  
\end{assumption}

\begin{theorem} \label{theorem3} 
Under Assumptions~\ref{assumption1} and \ref{assumption3}, 
for fixed load $\alpha>0$ the following limit 
\begin{equation} \label{QPSK_capacity_dif}
\frac{|C_{\mathrm{opt}}(\boldsymbol{Q}_{M};
\tilde{\gamma},\boldsymbol{G}_{\Delta_{\mathrm{t}},\Delta_{\mathrm{r}}})  
-C(\boldsymbol{Q}_{M};\tilde{\gamma},
\boldsymbol{G}_{\Delta_{\mathrm{t}},\Delta_{\mathrm{r}}})|}
{2\mathrm{min}\{L_{\mathrm{t}},L_{\mathrm{r}}\}}\to0
\end{equation}
holds uniformly for all receive antenna separations 
$\Delta_{\mathrm{r}}\in(0,1/2]$, covariance matrices 
$\boldsymbol{Q}_{M}\in\tilde{\mathfrak{S}}_{M}$, 
and all channel instances $\mathcal{C}\in\mathfrak{C}$ 
in the dense limit $\Delta_{\mathrm{t}}\to0$ after taking the large-system 
limit. 
\end{theorem}
\begin{IEEEproof}
See Section~\ref{sec5}.
\end{IEEEproof}

Assumption~\ref{assumption3} is a sufficient condition under which 
the linear minimum mean-square error (LMMSE) receiver with 
successive interference cancellation (SIC) operates in the low  
signal-to-interference-plus-noise ratio (SINR) regime for all stages. 
We shall present two examples of the precoding matrix $\boldsymbol{Q}_{M}$ 
satisfying Assumption~\ref{assumption3}. 

\begin{proposition} \label{proposition1} 
The identity precoding matrix 
$\boldsymbol{Q}_{M}^{1/2}=M^{-1/2}\boldsymbol{I}_{M}$ 
satisfies Assumption~\ref{assumption3}. 
\end{proposition}
\begin{IEEEproof}
It is straightforward to confirm $\boldsymbol{Q}_{M}\in\mathfrak{S}_{M}$ and 
$\|(2L_{\mathrm{t}})^{1/2}\boldsymbol{U}_{L_{\mathrm{t}},\Delta_{\mathrm{t}}}^{\mathrm{H}}
\boldsymbol{Q}_{M}^{1/2}\boldsymbol{e}_{M,m}\|^{2}=2\Delta_{\mathrm{t}}$
for all $m$. 
\end{IEEEproof}

Proposition~\ref{proposition1} is important for the case in which 
the transmitter has no ability to perform precoding. This situation may 
be realistic when low-quality amplifiers are used for all transmit antennas. 
Theorems~\ref{theorem1}--\ref{theorem3} imply that, when the true SNR 
$\gamma=4\Delta_{\mathrm{t}}\Delta_{\mathrm{r}}\tilde{\gamma}$ is considered, 
equal-power QPSK for $\Delta_{\mathrm{r}}=1/2$ achieves the normalized 
constrained capacity for the critically spaced case,  
\begin{IEEEeqnarray}{rl}
&\left(
 2\mathrm{min}\{L_{\mathrm{t}},L_{\mathrm{r}}\}
\right)^{-1}
C_{\mathrm{opt}}(M^{-1}\boldsymbol{I}_{M};
\gamma/(2\Delta_{\mathrm{t}}), 
\boldsymbol{G}_{\Delta_{\mathrm{t}},1/2}) 
\nonumber \\ 
=& \frac{1}{2\mathrm{min}\{L_{\mathrm{t}},L_{\mathrm{r}}\}}
\log\det\left(
 \boldsymbol{I} + 
 \frac{\gamma}{2L_{\mathrm{t}}}\boldsymbol{G}_{\Delta_{\mathrm{t}},1/2}
 \boldsymbol{G}_{\Delta_{\mathrm{t}},1/2}^{\mathrm{H}}
\right)
\nonumber \\ 
=& \frac{1}{2\mathrm{min}\{L_{\mathrm{t}},L_{\mathrm{r}}\}}
C_{\mathrm{opt}}((2L_{\mathrm{t}})^{-1}\boldsymbol{I}_{2L_{\mathrm{t}}}; 
\gamma,\boldsymbol{G}_{1/2,1/2})
\end{IEEEeqnarray}
in the dense limit $\Delta_{\mathrm{t}}\to0$ 
after taking the large-system limit. 

\begin{proposition} \label{proposition2}
For any $\boldsymbol{\Sigma}_{2L_{\mathrm{t}}+1}\in\mathfrak{S}_{2L_{\mathrm{t}}+1}$, 
the precoding matrix 
$\boldsymbol{Q}_{M}^{1/2}=\boldsymbol{U}_{L_{\mathrm{t}},\Delta_{\mathrm{t}}}
\boldsymbol{E}_{L_{\mathrm{t}}+1,M-(2L_{\mathrm{t}}+1)}
(\boldsymbol{\Sigma}_{2L_{\mathrm{t}}+1}^{1/2})
\boldsymbol{U}_{L_{\mathrm{t}},\Delta_{\mathrm{t}}}^{\mathrm{H}}$ 
satisfies Assumption~\ref{assumption3}. 
\end{proposition}
\begin{IEEEproof}
The condition 
$\boldsymbol{\Sigma}_{2L_{\mathrm{t}}+1}\in\mathfrak{S}_{2L_{\mathrm{t}}+1}$ 
implies $\boldsymbol{Q}_{M}\in\mathfrak{S}_{M}$. 
We shall prove $\boldsymbol{Q}_{M}\in\tilde{\mathfrak{S}}_{M}$. 
Let $\tilde{\boldsymbol{U}}_{L_{\mathrm{t}},\Delta_{\mathrm{t}}}$ denote the 
$M\times(2L_{\mathrm{t}}+1)$ matrix obtained by eliminating 
the columns $m\notin\mathcal{M}$ from 
$\boldsymbol{U}_{L_{\mathrm{t}},\Delta_{\mathrm{t}}}$. From the definition 
of $\boldsymbol{E}_{L_{\mathrm{t}}+1,M-(2L_{\mathrm{t}}+1)}
(\boldsymbol{\Sigma}_{2L_{\mathrm{t}}+1}^{1/2})$, 
\begin{IEEEeqnarray}{rl}
&\left\|
 (2L_{\mathrm{t}})^{1/2}\boldsymbol{Q}_{M}^{1/2}\boldsymbol{e}_{M,m}
\right\|^{2} 
\nonumber \\ 
=& 2L_{\mathrm{t}}
(\tilde{\boldsymbol{U}}_{L_{\mathrm{t}},\Delta_{\mathrm{t}}}^{\mathrm{H}}
\boldsymbol{e}_{M,m})^{\mathrm{H}}
\boldsymbol{\Sigma}_{2L_{\mathrm{t}}+1}
\tilde{\boldsymbol{U}}_{L_{\mathrm{t}},\Delta_{\mathrm{t}}}^{\mathrm{H}}
\boldsymbol{e}_{M,m}
\nonumber \\ 
\leq& \lambda_{\mathrm{max}}
\|\tilde{\boldsymbol{U}}_{L_{\mathrm{t}},\Delta_{\mathrm{t}}}^{\mathrm{H}}
\boldsymbol{e}_{M,m}\|^{2}, 
\end{IEEEeqnarray}
where the maximum eigenvalue 
$\lambda_{\mathrm{max}}>0$ of 
$2L_{\mathrm{t}}\boldsymbol{\Sigma}_{2L_{\mathrm{t}}+1}$ 
is uniformly bounded from Assumption~\ref{assumption2}. 
Since the $m$th column 
of $\boldsymbol{U}_{L_{\mathrm{t}},\Delta_{\mathrm{t}}}$ is equal to 
$\boldsymbol{s}_{L_{\mathrm{t}},\Delta_{\mathrm{t}}}(m/L_{\mathrm{t}})$ given by 
(\ref{signature}), we have 
\begin{equation} 
\|\tilde{\boldsymbol{U}}_{L_{\mathrm{t}},\Delta_{\mathrm{t}}}^{\mathrm{H}}
\boldsymbol{e}_{M,m}\|^{2}  
= \frac{\Delta_{\mathrm{t}}}{L_{\mathrm{t}}}\sum_{m'\in\mathcal{M}}
|e^{2\pi jm\Delta_{\mathrm{t}}m'/L_{\mathrm{t}}}|^{2} 
\leq 3\Delta_{\mathrm{t}}.  
\end{equation}
Thus, $\boldsymbol{Q}_{M}\in\tilde{\mathfrak{S}}_{M}$ holds. 
\end{IEEEproof}

Combining Theorems~\ref{theorem1}---\ref{theorem3}, 
from Proposition~\ref{proposition2} we can conclude that  
the precoded QPSK signaling achieves the normalized channel capacity 
for the critically spaced case in the dense limit $\Delta_{\mathrm{t}}\to0$ 
after taking the large-system limit, when one excludes the influence of 
the power gain obtained by spacing antenna elements densely. 
This result is analogous to the optimality of QPSK FTN signaling as 
the sampling period tends to zero~\cite{Yoo10}. 

\subsection{Discussion}
We shall make a comparison between the critically spaced case and 
the densely spaced case in terms of power consumption in the amplifiers. 
Since low peak-to-average power ratio (PAPR) results in low power 
consumption, PAPR is a key factor of the power consumption. 

For the critically spaced case, the precoded Gaussian signaling 
has to be used to achieve the constrained capacity. For the densely spaced 
case, on the other hand, the precoded QPSK scheme can be used from 
Theorem~\ref{theorem3}. 
For QPSK with no precoding in Proposition~\ref{proposition1} the 
instantaneous power is constant, so that the PAPR is the lowest, 
while the peak power of Gaussian signaling is unbounded. 
For the precoded QPSK scheme 
in Proposition~\ref{proposition2}, low PAPR is expected from the 
similarity between the precoding scheme and (localized) subcarrier mapping 
in single carrier frequency-division multiple access (SC-FDMA) 
systems~\cite{Myung06,Ochiai12}. See Section~\ref{sec6} for numerical 
comparisons between the precoded Gaussian and QPSK schemes. 

SC-FDMA systems are an alternate low-PAPR scheme of orthogonal 
frequency-division multiple-access (OFDMA) systems. In SC-FDMA systems, 
data symbols are converted to the frequency domain with the $m$-point DFT. 
After subcarrier mapping, the frequency-domain symbols are re-converted 
to the time domain via the $n$-point inverse DFT (IDFT). 
The precoding scheme in Proposition~\ref{proposition2} corresponds to 
the SC-FDMA scheme with the positions of the DFT and IDFT interchanged,   
if $\boldsymbol{\Sigma}_{2L_{\mathrm{t}}+1}$ is diagonal. The only difference is 
that the sizes of the two transforms are equal to each other in the precoding 
scheme, while $n>m$ holds in SC-FDMA systems. 

\section{Proof of Theorem~\ref{theorem3}}
\label{sec5} 
\subsection{Outline} 
The proof of Theorem~\ref{theorem3} consists of two steps. 
In the former step, a lower bound of the achievable 
rate~(\ref{QPSK_capacity}) is derived on the basis of the 
LMMSE-SIC~\cite{Li07,Takeuchi12,Takeuchi13}. 
After proving that the SINR in each stage of SIC tends to zero 
in the dense limit after taking the large-system limit, 
the interference-plus-noise is replaced with a CSCG random variable 
by using the fact that, when QPSK is used, the worst-case additive noise 
in the low SINR regime is Gaussian~\cite{Chan71}. 

In the latter step, we utilize the first-order optimality~\cite{Verdu02} of 
each data symbol for the AWGN channel in the low SINR regime to replace the 
data symbols by optimal Gaussian data symbols. Theorem~\ref{theorem3} follows 
from the optimality of the LMMSE-SIC for Gaussian signaling~\cite{Varanasi97}.  

\subsection{LMMSE-SIC}
We first use the chain rule~\cite{Cover06} for the mutual 
information~(\ref{QPSK_capacity}) to obtain 
\begin{equation} \label{QPSK_capacity1}
I(\boldsymbol{b};\boldsymbol{y}) 
= \sum_{m=0}^{M-1}I(b_{m};\boldsymbol{y}|b_{0},\ldots,b_{m-1}), 
\end{equation}
where $I(b_{m};\boldsymbol{y}|b_{0},\ldots,b_{m-1})$ corresponds to the 
achievable rate at stage~$m$ of SIC based on the optimal minimum 
mean-square error (MMSE) receiver. 
Consider the LMMSE estimator $\hat{b}_{m}$ of $b_{m}$ based on the known 
information $\boldsymbol{y}$ and $\{b_{m'}| m'\in[0:m)\}$. 
Since the LMMSE estimator is suboptimal, we have the lower bound 
\begin{equation} \label{LMMSE_Bound}
I(b_{m};\boldsymbol{y}|b_{0},\ldots,b_{m-1}) 
\geq I(b_{m};\hat{b}_{m}|b_{0},\ldots,b_{m-1}). 
\end{equation}

We shall derive the LMMSE estimator $\hat{b}_{m}$. 
From (\ref{effective_channel}), the output vector at stage~$m$ of SIC 
is given by 
\begin{equation} \label{SIC} 
\boldsymbol{y} - \sqrt{\tilde{\gamma}}\sum_{m'=0}^{m-1}\boldsymbol{a}_{m'}b_{m'} 
= \sqrt{\tilde{\gamma}}\sum_{m'=m}^{M-1}\boldsymbol{a}_{m'}b_{m'} 
+ \boldsymbol{w}. 
\end{equation}
In (\ref{SIC}), $\boldsymbol{a}_{m}\in\mathbb{C}^{N}$ denotes the $m$th 
column vector of the effective channel 
matrix~(\ref{effective_channel_matrix}). Thus, 
the LMMSE estimator $\hat{b}_{m}$ is given by \cite[Eq.~(8.66)]{Tse05}
\begin{equation} \label{LMMSE}
\hat{b}_{m} 
= \sqrt{\tilde{\gamma}}\boldsymbol{a}_{m}^{\mathrm{H}}\boldsymbol{\Xi}_{m}\left(
 \boldsymbol{y} - \sqrt{\tilde{\gamma}}\sum_{m'=0}^{m-1}\boldsymbol{a}_{m'}b_{m'} 
\right), 
\end{equation}
with 
\begin{equation} \label{MSE}
\boldsymbol{\Xi}_{m} 
= \left(
 \boldsymbol{I} + \tilde{\gamma}\sum_{m'=m+1}^{M-1}\boldsymbol{a}_{m'}
 \boldsymbol{a}_{m'}^{\mathrm{H}}
\right)^{-1}. 
\end{equation}

We derive the SINR for the LMMSE estimator $\hat{x}_{m}$. 
Define 
\begin{equation} \label{SINR} 
\rho_{m} 
= \tilde{\gamma}
\boldsymbol{a}_{m}^{\mathrm{H}}\boldsymbol{\Xi}_{m}\boldsymbol{a}_{m}. 
\end{equation}
Substituting (\ref{SIC}) into (\ref{LMMSE}) yields 
\begin{IEEEeqnarray}{rl} 
\frac{1}{\sqrt{\rho_{m}}}\hat{b}_{m} 
=& \sqrt{\rho_{m}}b_{m} + v_{m},
\label{LMMSE_output}
\end{IEEEeqnarray}
where the interference-plus-noise $v_{m}\in\mathbb{C}$ is given by 
\begin{equation}
\sqrt{\rho_{m}}v_{m} =
\tilde{\gamma}\sum_{m'=m+1}^{M-1}\boldsymbol{a}_{m}^{H}\boldsymbol{\Xi}_{m}
\boldsymbol{a}_{m'}b_{m'} 
+ \sqrt{\tilde{\gamma}}
\boldsymbol{a}_{m}^{\mathrm{H}}\boldsymbol{\Xi}_{m}\boldsymbol{w}. 
\end{equation}
Since the variance of $v_{m}$ is equal to $1$ from (\ref{MSE}), 
we find that $\rho_{m}$ is the SINR for the LMMSE estimator $\hat{b}_{m}$. 
Furthermore, (\ref{LMMSE_output}) implies that the lower 
bound~(\ref{LMMSE_Bound}) reduces to 
\begin{equation} \label{LMMSE_Bound2}
I(b_{m};\hat{b}_{m}|b_{0},\ldots,b_{m-1}) = I(b_{m};z_{m}),  
\end{equation} 
with $z_{m}=\sqrt{\rho_{m}}b_{m}+v_{m}$. 

\begin{lemma} \label{lemma4} 
Fix load $\alpha>0$. Under Assumptions~\ref{assumption1} 
and \ref{assumption3}, there exists some constant $A_{\alpha}>0$ such that 
the multiuser efficiency $\rho_{m}/(\Delta_{\mathrm{t}}\tilde{\gamma})$ 
normalized by $\Delta_{\mathrm{t}}$ is bounded from above by $A_{\alpha}$ 
for all $\Delta_{\mathrm{t}}, \Delta_{\mathrm{r}}\in(0,1/2]$, 
covariance matrices $\boldsymbol{Q}_{M}\in\tilde{\mathfrak{S}}_{M}$, 
channel instances $\mathcal{C}\in\mathfrak{C}$, and all $m\in[0: M)$. 
\end{lemma}
\begin{IEEEproof}
Since the maximum eigenvalue of (\ref{MSE}) is bounded from above by $1$,  
we have an upper bound for the SINR (\ref{SINR}), 
\begin{equation}
\frac{\rho_{m}}{\tilde{\gamma}} 
= \frac{\boldsymbol{a}_{m}^{\mathrm{H}}\boldsymbol{\Xi}_{m}
\boldsymbol{a}_{m}}{\|\boldsymbol{a}_{m}\|^{2}}
\|\boldsymbol{a}_{m}\|^{2}
< \left\|
 \boldsymbol{G}_{\Delta_{\mathrm{t}},\Delta_{\mathrm{r}}}
 \boldsymbol{U}_{L_{\mathrm{t}},\Delta_{\mathrm{t}}}^{\mathrm{H}}
 \boldsymbol{Q}_{M}^{1/2}\boldsymbol{e}_{M,m}
\right\|^{2}, 
\label{SINR_bound}
\end{equation}
where we have used the fact that $\boldsymbol{a}_{m}$ is the 
$m$th column of (\ref{effective_channel_matrix}). 
Repeating the same argument yields 
\begin{equation}
\frac{\rho_{m}}{\tilde{\gamma}} 
< \sigma_{\mathrm{max}}^{2}
\left\|
 (2L_{\mathrm{t}})^{1/2}\boldsymbol{Q}_{M}^{1/2}\boldsymbol{e}_{M,m}
\right\|^{2}, 
\end{equation}
where $\sigma_{\mathrm{max}}>0$ denotes the maximum singular value of the 
channel matrix 
$(2L_{\mathrm{t}})^{-1/2}\boldsymbol{G}_{\Delta_{\mathrm{t}},\Delta_{\mathrm{r}}}$. 
From Assumptions~\ref{assumption1} and \ref{assumption3}, we find that 
Lemma~\ref{lemma4} holds. 
\end{IEEEproof}

A further lower bound for (\ref{LMMSE_Bound2}) is derived by using the fact 
that Gaussian noise is the worst-case additive noise for the real AWGN 
channel with binary phase-shift keying (BPSK)~\cite{Chan71}. 
Since the real and imaginary parts $\Re[b_{m}]$ and $\Im[b_{m}]$ are 
independent for QPSK, (\ref{LMMSE_Bound2}) reduces to 
\begin{equation}
I(b_{m};z_{m}) 
= H(\Re[b_{m}]) + H(\Im[b_{m}]) - H(b_{m}|z_{m}). 
\end{equation}
Using the chain rule for entropy~\cite{Cover06} yields 
\begin{IEEEeqnarray}{rl}
H(b_{m}|z_{m}) 
=& H(\Re[b_{m}]|z_{m}) + H(\Im[b_{m}]|z_{m}, \Re[b_{m}]) 
\nonumber \\ 
\leq& H(\Re[b_{m}]|\Re[z_{m}]) + H(\Im[b_{m}]|\Im[z_{m}]), 
\end{IEEEeqnarray}
where the inequality follows from the fact that conditioning reduces 
entropy~\cite{Cover06}. Thus, we have 
\begin{equation}
I(b_{m};z_{m}) 
\geq I(\Re[b_{m}];\Re[z_{m}]) + I(\Im[b_{m}];\Im[z_{m}]). 
\end{equation}
Replacing $\Re[v_{m}]$ and $\Im[v_{m}]$ in $z_{m}$ with the worst-case additive 
noise for the BPSK AWGN channel in the low SNR regime~\cite{Chan71}, 
we arrive at the lower bound, 
\begin{equation} \label{worst_case}
I(b_{m};z_{m}) \geq I(b_{m};\sqrt{\rho_{m}}b_{m}+v_{m}^{\mathrm{G}}), 
\end{equation}
with $v_{m}^{\mathrm{G}}\sim\mathcal{CN}(0,1)$, in the dense limit 
$\Delta_{\mathrm{t}}\to0$ after taking the large-system limit 
uniformly for all $\Delta_{\mathrm{r}}\in(0,1/2]$, covariance matrices 
$\boldsymbol{Q}_{M}\in\tilde{\mathfrak{S}}_{M}$, and all 
channel instances $\mathcal{C}\in\mathfrak{C}$.  
Applying (\ref{LMMSE_Bound}), (\ref{LMMSE_Bound2}), and (\ref{worst_case}) to 
(\ref{QPSK_capacity1}) yields 
\begin{equation} \label{QPSK_capacity2}
\frac{1}{M}I(\boldsymbol{b};\boldsymbol{y}) 
\geq \frac{1}{M}\sum_{m=0}^{M-1}I(b_{m};\sqrt{\rho_{m}}b_{m} + v_{m}^{\mathrm{G}})
\end{equation} 
in the dense limit after taking the large-system limit. 

\begin{remark}
One may attempt to use the central limit theorem in order to replace 
the interference-plus-noise $v_{m}$ by the CSCG random variable 
$v_{m}^{\mathrm{G}}$in the large-system limit. The asymptotic Gaussianity of 
the interference was proved in \cite{Guo02} when the channel matrix 
$\boldsymbol{H}$ has i.i.d.\ elements. However, it is not clear whether 
there exists a pathological channel instance $\mathcal{C}\in\mathfrak{C}$ 
such that $v_{m}$ is non-Gaussian even in the large-system limit. 
If $v_{m}$ converges to $v_{m}^{\mathrm{G}}$, we could postulate quadrature 
amplitude modulation (QAM) data symbols to obtain Theorem~\ref{theorem3}, 
instead of QPSK. 
\end{remark}

\subsection{Optimality of LMMSE-SIC} 
We use the first-order optimality of $b_{m}$ to evaluate the mutual 
information $I(b_{m};\sqrt{\rho_{m}}b_{m}+v_{m}^{\mathrm{G}})$. 
Let $b_{m}^{\mathrm{G}}\sim\mathcal{CN}(0,1)$ denote a CSCG data symbol 
with unit power. Since any zero-mean and unit-power signaling is first-order 
optimal for the AWGN channel~\cite[Theorem~4]{Verdu02}, we have 
\begin{equation} \label{SO_optimality}
\frac{
 |I(b_{m};\sqrt{\rho_{m}}b_{m}+v_{m}^{\mathrm{G}}) 
 - I(b_{m}^{\mathrm{G}};\sqrt{\rho_{m}}b_{m}^{\mathrm{G}}+v_{m}^{\mathrm{G}})|
}{\rho_{m}} \to 0
\end{equation} 
uniformly for all $\Delta_{\mathrm{r}}\in(0,1/2]$, 
covariance matrices $\boldsymbol{Q}_{M}\in\tilde{\mathfrak{S}}_{M}$, 
and all channel instances $\mathcal{C}\in\mathfrak{C}$ in the dense limit 
$\Delta_{\mathrm{t}}\to0$ after taking the large-system limit. 

Applying (\ref{SO_optimality}) to (\ref{QPSK_capacity2}), 
from Lemma~\ref{lemma4} we find 
\begin{equation}
\frac{1}{M}I(\boldsymbol{b};\boldsymbol{y}) 
\geq \frac{1}{M}C_{\mathrm{opt}}(\boldsymbol{Q}_{M};
\tilde{\gamma}, \boldsymbol{G}_{\Delta_{\mathrm{t}},\Delta_{\mathrm{r}}}) 
+ o(\Delta_{\mathrm{t}}),  
\label{QPSK_capacity3}
\end{equation}
in the dense limit $\Delta_{\mathrm{t}}\to0$ after taking the large-system 
limit. In the derivation of (\ref{QPSK_capacity3}), 
we have used the fact that the LMMSE-SIC for Gaussian signaling achieves 
the constrained capacity 
$C_{\mathrm{opt}}(\boldsymbol{Q}_{M};
\tilde{\gamma}, \boldsymbol{G}_{\Delta_{\mathrm{t}},\Delta_{\mathrm{r}}})$ 
given by (\ref{capacity})~\cite{Varanasi97}. 
Dividing both sides by $\Delta_{\mathrm{t}}$, we have 
\begin{equation} \label{QPSK_capacity4}
\frac{L_{\mathrm{t}}}{2\mathrm{min}\{L_{\mathrm{t}},L_{\mathrm{r}}\}}
\frac{I(\boldsymbol{b};\boldsymbol{y}) 
- C_{\mathrm{opt}}(\boldsymbol{Q}_{M};
\tilde{\gamma}, \boldsymbol{G}_{\Delta_{\mathrm{t}},\Delta_{\mathrm{r}}})}
{L_{\mathrm{t}}}\geq0
\end{equation}
in the dense limit $\Delta_{\mathrm{t}}\to0$ after taking the large-system 
limit. Since the upper bound 
$C_{\mathrm{opt}}(\boldsymbol{Q}_{M};
\tilde{\gamma}, \boldsymbol{G}_{\Delta_{\mathrm{t}},\Delta_{\mathrm{r}}})
-I(\boldsymbol{b};\boldsymbol{y})\geq0$ is trivial, we arrive at 
Theorem~\ref{theorem3}. 

\section{Numerical Results} \label{sec6}
\subsection{Simulation Conditions}
In all numerical results, we assume the $P$-path Rayleigh fading model 
as an example of the channel matrix 
$\boldsymbol{G}_{\Delta_{\mathrm{t}},\Delta_{\mathrm{r}}}$ in the angular 
domain---given by (\ref{channel_gain}),   
\begin{IEEEeqnarray}{rl}
(\boldsymbol{G}_{\Delta_{\mathrm{t}},\Delta_{\mathrm{r}}})_{n,m}
= \sqrt{4L_{\mathrm{t}}L_{\mathrm{r}}}
\sum_{p=0}^{P-1}&a_{p}f_{L_{\mathrm{r}},\Delta_{\mathrm{r}}}\left(
 \frac{n}{L_{\mathrm{r}}} - \cos\phi_{\mathrm{r},p}
\right)
\nonumber \\ 
\cdot& f_{L_{\mathrm{t}},\Delta_{\mathrm{t}}}^{*}\left(
 \frac{m}{L_{\mathrm{t}}} - \cos\phi_{\mathrm{t},p}
\right). \label{Rayleigh}
\end{IEEEeqnarray}
In (\ref{Rayleigh}), the attenuation $\{a_{p}| p\in[0:P)\}$ are independently 
sampled from the CSCG distribution with variance $1/P$. Furthermore, 
the angles $\{\phi_{\mathrm{t},p}, \phi_{\mathrm{r},p}|p\in[0:P)\}$ of 
departure and incident are independently drawn from the uniform 
distribution on $[0, 2\pi)$. The condition 
$P\geq\mathrm{min}\{2L_{\mathrm{t}},2L_{\mathrm{r}}\}$ is necessary 
for achieving the full spatial degrees of freedom. 

We focus on single-user massive MIMO downlink, in which the transmitter is 
a base station with a large antenna array, while the receiver corresponds 
to a user with a small antenna array. Thus, we consider the case in which 
the normalized length $L_{\mathrm{t}}$ of the transmit antenna array is 
larger than the normalized length $L_{\mathrm{r}}$ of the receive antenna 
array. The transmitter may equip densely spaced antennas 
($\Delta_{\mathrm{t}}\leq1/2$), whereas the receiver has critically spaced 
antennas ($\Delta_{\mathrm{r}}=1/2$). Thus, the number 
$M=L_{\mathrm{t}}/\Delta_{\mathrm{t}}$ of transmit antennas may be much larger 
than the number $N=L_{\mathrm{r}}/\Delta_{\mathrm{r}}$ of receive antennas. 

Consider the precoding scheme 
$\boldsymbol{x}=\boldsymbol{Q}_{M}^{1/2}\boldsymbol{b}$, with 
$\boldsymbol{Q}_{M}=\boldsymbol{U}_{L_{\mathrm{t}},\Delta_{\mathrm{t}}}
\boldsymbol{\Sigma}_{M}\boldsymbol{U}_{L_{\mathrm{t}},\Delta_{\mathrm{t}}}^{\mathrm{H}}$. 
In estimating the constrained capacity 
($\boldsymbol{b}\sim\mathcal{CN}(\boldsymbol{0},\boldsymbol{I})$), 
we consider the ergodic constrained 
capacity averaged over sufficiently many independent channel instances, 
called constrained capacity simply. For the case of full channel state 
information (CSI), the input covariance matrix 
$\boldsymbol{\Sigma}_{M}\in\mathfrak{S}_{M}$ was optimized with 
the water-filling algorithm~\cite{Tse05}. For the case of CSI at the receiver 
(CSIR), we considered the two diagonal input covariance matrices 
$\boldsymbol{\Sigma}_{2L_{\mathrm{t}}}=(2L_{\mathrm{t}})^{-1}
\boldsymbol{I}_{2L_{\mathrm{t}}}$ for the critically spaced case 
$\Delta_{\mathrm{t}}=1/2$ and 
$\boldsymbol{\Sigma}_{M}=(2L_{\mathrm{t}}+1)^{-1}
\mathrm{diag}\{\boldsymbol{1}_{L_{\mathrm{t}}+1}, \boldsymbol{0}, 
\boldsymbol{1}_{L_{\mathrm{t}}}\}$ for the densely spaced case 
$\Delta_{\mathrm{t}}<1/2$, in which $\boldsymbol{1}_{n}$ denotes the 
$n$-dimensional vector whose elements are all one. In the densely spaced 
case, the power is uniformly allocated to the antenna indices in which 
the transmit beamforming patterns have main lobes. 

\subsection{Precoded Gaussian Signaling}
Consider the case of Gaussian signaling. We first investigate 
the influence of channel gains for indices 
$m\in\mathcal{M}^{\mathrm{c}}=(L_{\mathrm{t}}:M-L_{\mathrm{t}})$, in which 
the transmit beamforming patterns have no main lobes, as shown in 
Fig.~\ref{fig1}. 

\begin{figure}[t]
\begin{center}
\includegraphics[width=\hsize]{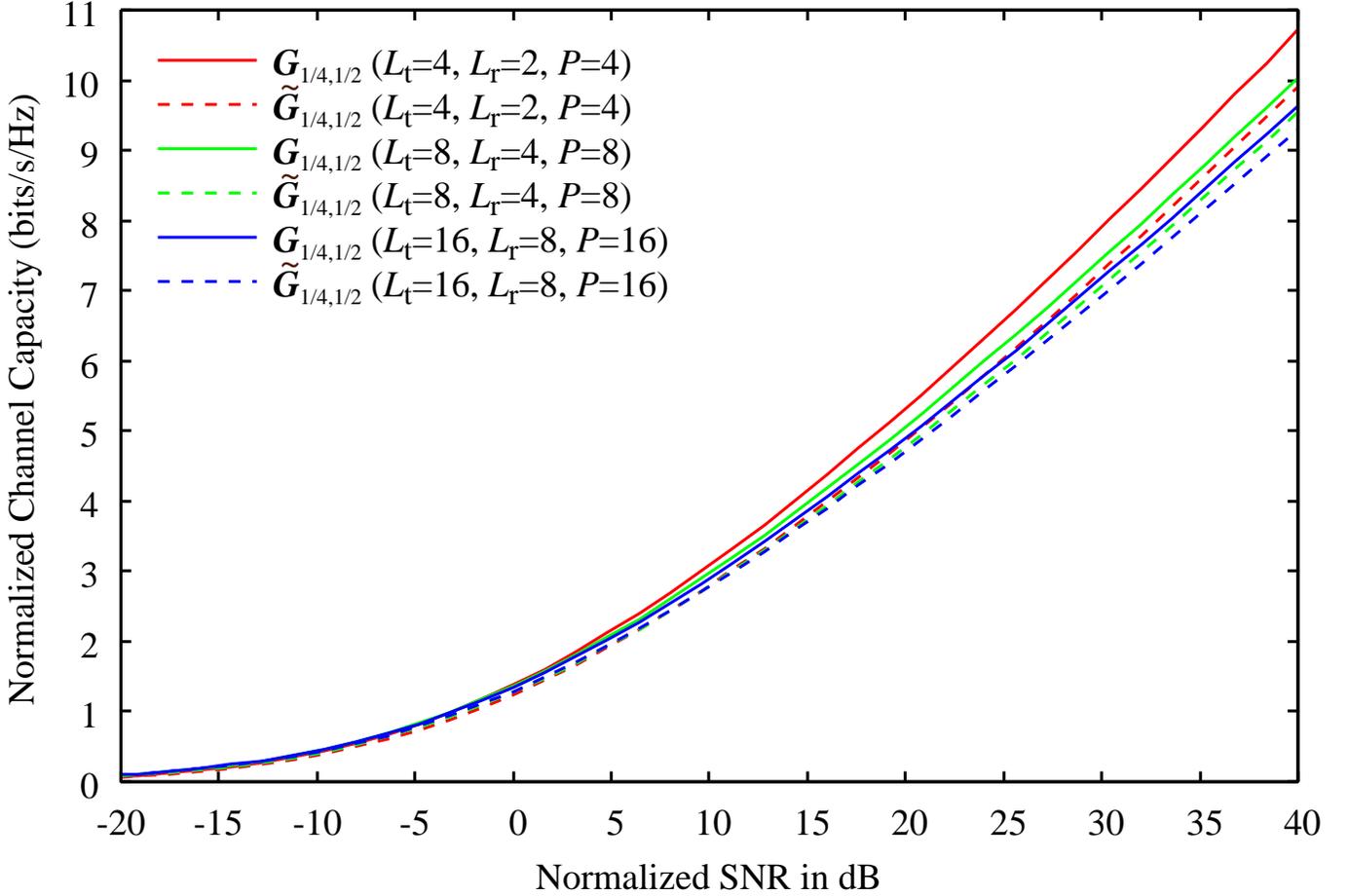}
\caption{
Normalized channel capacity $C_{\mathrm{opt}}/N$ versus normalized SNR 
$\tilde{\gamma}$ under the full CSI assumption for 
$\Delta_{\mathrm{t}}=1/4$ and $\Delta_{\mathrm{r}}=1/2$. 
}
\label{fig2}
\end{center}
\end{figure}

Figure~\ref{fig2} shows the normalized channel capacities  
for the two channel matrices $\boldsymbol{G}_{\Delta_{\mathrm{t}},1/2}$ and 
$\tilde{\boldsymbol{G}}_{\Delta_{\mathrm{t}},1/2}$ given by (\ref{hat_G}) 
under the assumption of full channel state information (CSI). 
We find that there are gaps between the two capacities at high SNRs. 
The gap decreases slowly as $L_{\mathrm{t}}$ and $L_{\mathrm{r}}$ grows at 
the same rate. This result is consistent\footnote{
Numerical simulations showed that the average of the individual 
differences for identical channel instances decreased slowly, although 
no figure is presented.} with Theorem~\ref{theorem1}. 

\begin{figure}[t]
\begin{center}
\includegraphics[width=\hsize]{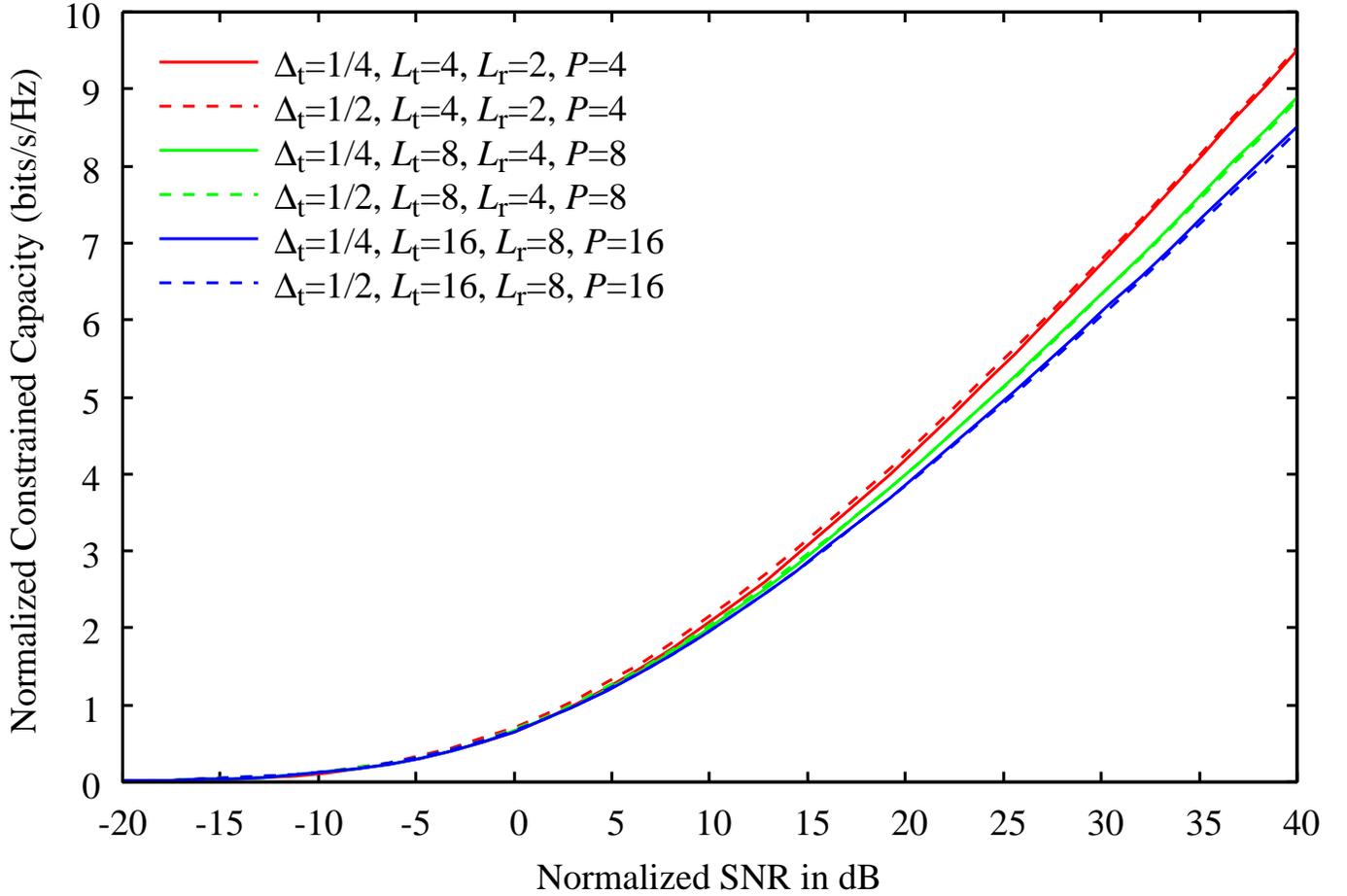}
\caption{
Normalized constrained capacity $C_{\mathrm{opt}}/N$ versus normalized 
SNR~$\tilde{\gamma}$ under the CSIR assumption for $\Delta_{\mathrm{r}}=1/2$. 
}
\label{fig3}
\end{center}
\end{figure}

We next make comparisons between the critically spaced case 
$\Delta_{\mathrm{t}}=1/2$ and the densely spaced case $\Delta_{\mathrm{t}}<1/2$. 
Figure~\ref{fig3} shows the constrained capacities for the two channel 
matrices $\boldsymbol{G}_{1/2,1/2}$ and $\boldsymbol{G}_{\Delta_{\mathrm{t}},1/2}$ 
under the CSIR assumption. We find that the two capacities are 
indistinguishable from each other even for small systems. This result 
implies that the difference~(\ref{capacity_difference}) in 
Theorem~\ref{theorem2} converges very quickly on average, while the proof 
of Theorem~\ref{theorem2} predicts slow convergence for the worst channel 
instance. 

\subsection{Precoded Non-Gaussian Signaling}

\section{Conclusion} \label{sec7}

\appendices

\section{Proof of Lemma~\ref{lemma2}} \label{proof_lemma2}
In proving Lemma~\ref{lemma2}, we use the following lemma. 
\begin{lemma} \label{lemma_appen0}
\begin{equation}
\sum_{k=n+1}^{\infty}\frac{1}{k^{2}}=O(n^{-1}) 
\quad \hbox{as $n\to\infty$.}
\end{equation}
\end{lemma}
\begin{IEEEproof}
The lemma follows from the bound 
\begin{equation} \label{Basel_prob}
\frac{\pi^{2}}{6}\frac{2n(2n-1)}{(2n+1)^{2}} 
< \sum_{k=1}^{n}\frac{1}{k^{2}} 
< \frac{\pi^{2}}{6}\frac{2n(2n+2)}{(2n+1)^{2}}. 
\end{equation}

We review a classical elementary proof of (\ref{Basel_prob}). 
Start with the identity for $2n+1\in\mathbb{N}$ 
\begin{equation}
\frac{e^{j(2n+1)x}}{(\sin x)^{2n+1}} 
= \left(
 \frac{1}{\tan x}+j
\right)^{2n+1}. 
\end{equation}
Binomial-expanding the right-hand side (RHS) and then comparing the 
imaginary parts on both sides, we obtain 
\begin{equation} \label{binom_identity} 
\frac{\sin\{(2n+1)x\}}{(\sin x)^{2n+1}} 
= \sum_{r=0}^{n}
\binom{2n+1}{2r+1}(-1)^{r}t^{n-r}, 
\end{equation} 
with $t=(\tan x)^{-2}$. 

Consider $x_{k}=k\pi/(2n+1)$ for $k=1,\ldots,n$. 
It is straightforward to confirm that $\{(\tan x_{k})^{-2}| k\in[1: n]\}$ are 
the roots of the $n$th-degree polynomial on the RHS of 
(\ref{binom_identity}) with respect to $t$. From Vieta's formulas, we have 
\begin{equation} \label{identity1}
\sum_{k=1}^{n}\frac{1}{\tan^{2}x_{k}} 
= \binom{2n+1}{1}^{-1}\binom{2n+1}{3} 
= \frac{2n(2n-1)}{6}. 
\end{equation}
Applying the identity $(\sin x_{k})^{-2}=1+(\tan x_{k})^{-2}$ yields 
\begin{equation} \label{identity2}
\sum_{k=1}^{n}\frac{1}{\sin^{2}x_{k}}  
= \frac{2n(2n+2)}{6}. 
\end{equation}
The bound~(\ref{Basel_prob}) follows from (\ref{identity1}), 
(\ref{identity2}), and the bound $\sin^{2}x_{k}<x_{k}^{2}<\tan^{2}x_{k}$ 
for all $x_{k}=k\pi/(2n+1)$. 
\end{IEEEproof}

Let us prove Lemma~\ref{lemma2}. 
From $\mathcal{K}_{2}=\emptyset$ for $\Delta=1/2$, by definition 
$I_{2}(\Omega,\Omega';L,\Delta)=0$ holds. Thus, we assume $\Delta<1/2$. 
For notational simplicity, we write $f_{L,\Delta}$ as $f$. 

Using the Cauchy-Schwarz inequality, we have an upper-bound for 
$|I_{2}(\Omega,\Omega';L,\Delta)|^{2}$ given by (\ref{partial_sum}), 
\begin{equation}
|I_{2}(\Omega,\Omega';L,\Delta)|^{2} \leq F(\Omega)F(\Omega'), 
\end{equation}
with 
\begin{equation}  
F(\Omega) 
= 
\sum_{k=L+1}^{N-L-1}\left|
 f\left(
  \frac{k}{L} - \Omega
 \right)
\right|^{2}. 
\end{equation}
Thus, it is sufficient to prove that $s_{L}F(\Omega)$ is uniformly 
bounded for all $\Omega\in\mathcal{D}_{L}$ and $\Delta\in(0,1/2)$. 

Since $|f(\Omega)|$ is an even periodic function with 
period $N/L$ from Property~\ref{property1}, 
i.e.\ $|f(\Omega)|=|f(N/L-\Omega)|$, we use the property 
$N=\lceil N/2\rceil+\lfloor N/2\rfloor$ to represent $F(\Omega)$ as 
\begin{IEEEeqnarray}{rl}
F(\Omega) 
=& \sum_{k=L+1}^{\lceil N/2\rceil-1}
\left|
 f\left(
  \frac{k}{L} - \Omega
 \right)
\right|^{2}
+ \sum_{k=L+1}^{\lfloor N/2\rfloor}
\left|
 f\left(
  \frac{k}{L} + \Omega
 \right)
\right|^{2}
\nonumber \\
\leq&\sum_{k=L+1}^{\lfloor N/2\rfloor}\left\{
 \left|
  f\left(
   \frac{k}{L} - \Omega
  \right)
 \right|^{2}
 + \left|
  f\left(
   \frac{k}{L} + \Omega
  \right)
 \right|^{2}
\right\}. \label{F_upper_bound}
\end{IEEEeqnarray}
Since the upper bound~(\ref{F_upper_bound}) is an even function of $\Omega$, 
without loss of generality, we assume $\Omega\in[0, 1-s_{L}/L]$. 
In order to use the upper bound~(\ref{sinc_bound}), we 
decompose (\ref{F_upper_bound}) into two terms,   
\begin{equation} \label{F_bound}
F(\Omega) \leq F_{1}(\Omega) + F_{2}(\Omega), 
\end{equation}
with 
\begin{equation}
F_{1} 
= \sum_{k=L+1}^{\lfloor N/2\rfloor}
\left|
 f\left(
  \frac{k}{L} - \Omega
 \right)
\right|^{2}
+ \sum_{k=L+1}^{\lceil N/2 - L\Omega\rceil-1}
\left|
 f\left(
  \frac{k}{L} + \Omega
 \right)
\right|^{2},
\end{equation}
\begin{equation}
F_{2}  
= \sum_{k=\lceil N/2 - L\Omega\rceil}^{\lfloor N/2\rfloor}
\left|
 f\left(
  \frac{k}{L} + \Omega
 \right)
\right|^{2}. 
\end{equation}
Note that $k+L\Omega\geq N/2$ holds in $F_{2}(\Omega)$ 
when $k$ runs from $\lceil N/2 -L\Omega \rceil$. 

We first upper-bound $F_{1}(\Omega)$. Substituting the 
expression~(\ref{func_f1}) and using the upper bound~(\ref{sinc_bound}), 
we find 
\begin{IEEEeqnarray}{rl}
F_{1}(\Omega) 
<& \frac{1}{4}\sum_{k=L+1}^{\lfloor N/2\rfloor}\frac{1}{(k-L\Omega)^{2}} 
+ \frac{1}{4}\sum_{k=L+1}^{\lceil N/2-L\Omega\rceil-1}\frac{1}{(k+L\Omega)^{2}} 
\nonumber \\ 
<& \frac{1}{4}\sum_{k=s_{L}+1}^{\infty}\frac{1}{k^{2}}
+ \frac{1}{4}\sum_{k=L+1}^{\infty}\frac{1}{k^{2}},   
\label{F1_bound}
\end{IEEEeqnarray}
for all $\Omega\in[0,1-s_{L}/L]$. 

We next evaluate $F_{2}(\Omega)$. Property~\ref{property1} implies
the symmetry $|f(\Omega)|=|f(\Omega-1/\Delta)|=|f(1/\Delta-\Omega)|$. 
From $k+L\Omega\geq N/2$ for $k\geq\lceil N/2 -L\Omega \rceil$, 
we use the upper bound~(\ref{sinc_bound}) to obtain  
\begin{IEEEeqnarray}{rl}
&F_{2}(\Omega) 
\nonumber \\  
=& \sum_{k=\lceil N/2 -L\Omega \rceil}^{\lfloor N/2\rfloor}
|\mathrm{sinc}_{N}(N-k-L\Omega)|^{2} 
\nonumber \\ 
<& \frac{1}{4}\sum_{k=\lceil N/2 -L\Omega \rceil}^{\lfloor N/2\rfloor-1}
\frac{1}{(N-k-L\Omega)^{2}}
+ |\mathrm{sinc}_{N}(\lceil N/2\rceil-L\Omega)|
\nonumber \\ 
<& \frac{1}{4}\sum_{k=\lceil N/2\rceil-L+s_{L}}^{\infty}
\frac{1}{k^{2}}
+\frac{1}{4\lfloor N/2\rfloor^{2}}, \label{F2_bound}
\end{IEEEeqnarray}
for all $\Omega\in[0,1-s_{L}/L]$. In the derivation of the last inequality, 
we have used the following upper bounds: 
\begin{IEEEeqnarray}{rl}
|\mathrm{sinc}_{N}(\lceil N/2\rceil-L\Omega)|
\leq& \frac{1}{2|\lceil N/2\rceil-L\Omega|} 
\nonumber \\ 
\leq& \frac{1}{2|\lceil N/2\rceil-L+s_{L}|},  
\label{first_bound}
\end{IEEEeqnarray}
for all $\lceil N/2\rceil-L\Omega\leq N/2$. Otherwise, 
\begin{equation}
|\mathrm{sinc}_{N}(\lceil N/2\rceil-L\Omega)|
= |\mathrm{sinc}_{N}(\lfloor N/2\rfloor+L\Omega)| 
\leq \frac{1}{2\lfloor N/2\rfloor}.
\end{equation}

From Lemma~\ref{lemma_appen0}, (\ref{F_bound}), (\ref{F1_bound}), 
and (\ref{F2_bound}), 
we find that $s_{L}F(\Omega)$ is uniformly bounded for all 
$\Omega\in\mathcal{D}_{L}$ and $\Delta\in(0,1/2)$ 
in the limit $L, N\to\infty$ with $\Delta=L/N$ fixed.

\section{Proof of Lemma~\ref{lemma3}}
\label{proof_lemma3}
For notational convenience, $D_{k,k}(\Omega)$ is simply written as 
$D_{k}(\Omega)$. We first prove the following lemma. 
\begin{lemma} \label{lemma_appen1}
There exists some constant $A>0$ such that 
\begin{equation} \label{D_k_bound} 
|D_{k}(\Omega)|^{2} 
< \frac{1}{4L^{2}} + \frac{A}{(2L-|k-L\Omega|)^{2}}, 
\end{equation}
for all $k\in[0:L]$, $\Omega\in\mathcal{D}_{L}$, and all $\Delta\in(0,1/2]$. 
\end{lemma}
\begin{IEEEproof}
Let $x=k-L\Omega\in[-L+s_{L}, 2L-s_{L}]$. 
From (\ref{func_f1}), (\ref{sinc}), and (\ref{D_k0}), we have  
\begin{equation} \label{D_k}
|D_{k}(\Omega)|^{2} 
= \sin^{2}(\pi x)\left\{
 |d_{\Delta}(x)|^{2}
 + \frac{1}{L^{2}}\left(
  \frac{1}{2} - \Delta
 \right)^{2}
\right\}, 
\end{equation}
with 
\begin{equation}
d_{\Delta}(x) 
= \frac{1}{L}\left\{
 \frac{\Delta}{\tan(\pi\Delta x/L)} - \frac{1}{2\tan(\pi x/(2L))}
\right\}, 
\end{equation} 
where we define $d_{\Delta}(0)=\lim_{x\to0}d_{\Delta}(x)=0$. 

The upper bound~(\ref{D_k_bound}) follows from (\ref{D_k}) and the 
following bound:  
\begin{equation} \label{d_bound} 
|d_{\Delta}(x)| < \bar{d}(x)=\frac{\sqrt{A}}{2L-|x|}, 
\quad \hbox{for $x\in(-2L, 2L)$.} 
\end{equation}
for some constant $A>0$. 

Let us prove the upper bound~(\ref{d_bound}).  
Since $|d_{\Delta}(x)|$ is an even function, we only consider the interval 
$[0, 2L)$. It is straightforward to confirm $d_{\Delta}(x)\geq0$ for 
$x\in[0, 2L)$ and $\Delta\in(0,1/2]$, because $u/\tan u$ is monotonically 
decreasing for $u\in[0, \pi)$. Furthermore, 
$d_{\Delta}(x)\leq d_{\Delta'}(x)$ for $\Delta'\leq \Delta$ holds. 
Taking the limit $\Delta\to0$, we have the upper bound 
$d_{\Delta}(x)\leq d_{0}(x)$, with 
\begin{equation}
d_{0}(x) = \frac{1}{\pi x} - \frac{1}{2L\tan(\pi x/(2L))}. 
\end{equation} 

Let $y=\pi x/(2L)\in[0, \pi)$. 
Evaluating the product $\pi(2L-x)d_{\Delta}(x)$ yields 
\begin{equation}
\pi(2L-x)d_{\Delta}(x) 
\leq \frac{\pi -y}{y} - \frac{\pi -y}{\tan y} 
\equiv \tilde{d}_{0}(y). 
\end{equation}
The boundedness of $\tilde{d}_{0}(y)$ follows from 
$\lim_{y\to0}\tilde{d}_{0}(y)=0$, 
$\lim_{y\to\pi}\tilde{d}_{0}(y)=1$, and the continuity of $\tilde{d}_{0}(y)$ 
on $(0,\pi)$. Thus, the upper bound~(\ref{d_bound}) holds. 
\end{IEEEproof}

Lemma~\ref{lemma3} can be proved in the same manner as in the proof of 
Lemma~\ref{lemma2}, by using Lemma~\ref{lemma_appen1}   
instead of (\ref{sinc_bound}). 
Using the Cauchy-Schwarz inequality, we upper-bound the quantity 
$|I_{3}(\Omega,\Omega';L,\Delta)|^{2}$ given by (\ref{difference_sum}) as  
\begin{equation}
|I_{3}(\Omega,\Omega';L,\Delta)|^{2}\leq G(\Omega)G(\Omega'), 
\end{equation}
with 
\begin{equation}
G(\Omega) 
= \sum_{k=0}^{2L-1}|D_{k,\kappa_{L,\Delta}(k)}(\Omega)|^{2} 
+ |f_{L,\Delta}(1-\Omega)|^{2}. 
\end{equation}
Thus, it is sufficient to prove that $s_{L}G(\Omega)$ is uniformly bounded 
for all $\Omega\in\mathcal{D}_{L}$ and $\Delta\in(0,1/2]$. 

From (\ref{func_f1}), (\ref{kappa}), (\ref{D_k0}), and 
the first two properties in Property~\ref{property1}, we have 
\begin{IEEEeqnarray}{rl} 
G(\Omega) 
=& \sum_{k=0}^{L-1}|D_{k}(\Omega)|^{2}
+ \sum_{k=1}^{L}|D_{k}(-\Omega)|^{2} 
\nonumber \\ 
&+ |\mathrm{sinc}_{L/\Delta}(L(1-\Omega))|^{2}. 
\label{G_upper_bound}
\end{IEEEeqnarray}
Applying the upper bound~(\ref{D_k_bound}) to (\ref{G_upper_bound}) yields 
\begin{equation} \label{G_upper_bound2}
G(\Omega) < 
\frac{1}{2L} + AS(\Omega)
+ |\mathrm{sinc}_{L/\Delta}(L(1-\Omega))|^{2},     
\end{equation} 
with 
\begin{equation} \label{S_sum}
S(\Omega) = \sum_{k=0}^{L}\left\{
 \frac{1}{(2L-|k-L\Omega|)^{2}} 
 + \frac{1}{(2L-|k+L\Omega|)^{2}} 
\right\}. 
\end{equation}

We first evaluate $|\mathrm{sinc}_{L/\Delta}(L(1-\Omega))|$. 
When $L(1-\Omega)\leq N/2$ holds, from the upper bound~(\ref{sinc_bound}) 
we have 
\begin{equation}
|\mathrm{sinc}_{L/\Delta}(L(1-\Omega))| 
\leq \frac{1}{2L(1-\Omega)}
\leq \frac{1}{2s_{L}}, 
\end{equation}
for all $\Omega\in\mathcal{D}_{L}$. Otherwise, 
\begin{IEEEeqnarray}{rl}
|\mathrm{sinc}_{L/\Delta}(L(1-\Omega))| 
=&|\mathrm{sinc}_{L/\Delta}(N-L(1-\Omega))|
\nonumber \\ 
\leq& \frac{1}{2(N-2L+s_{L})}. 
\end{IEEEeqnarray}
Combining the two upper bounds yields 
\begin{equation} \label{sinc_bound2}
|\mathrm{sinc}_{L/\Delta}(L(1-\Omega))| 
\leq \frac{1}{2s_{L}}, 
\end{equation}
for all $\Omega\in\mathcal{D}_{L}$ and $\Delta\in(0,1/2]$. 

We next evaluate (\ref{S_sum}). 
Since $S(\Omega)$ is an even function of $\Omega\in\mathcal{D}_{L}$, without 
loss of generality, $\Omega\in[0,1-s_{L}/L]$ is assumed. 
We decompose $S(\Omega)$ into the sum $S_{1}(\Omega)+S_{2}(\Omega)$, 
with
\begin{equation}
S_{1}(\Omega)
= \sum_{k=0}^{\lfloor L\Omega\rfloor}\frac{1}{(2L+k-L\Omega)^{2}}
+ \sum_{k=0}^{L}\frac{1}{\{2L-(k+L\Omega)\}^{2}},  
\end{equation}
\begin{equation}
S_{2}(\Omega)
= \sum_{k=\lfloor L\Omega\rfloor+1}^{L}
\frac{1}{\{2L-(k-L\Omega)\}^{2}}. 
\end{equation}
For $S_{1}(\Omega)$, we have 
\begin{IEEEeqnarray}{rl} 
S_{1}(\Omega) 
\leq& \sum_{k=0}^{\lfloor L\Omega\rfloor}\frac{1}{(L+s_{L}+k)^{2}} 
+ \sum_{k=0}^{L}\frac{1}{(L+s_{L}-k)^{2}} 
\nonumber \\ 
<& \sum_{k=L+s_{L}}^{\infty}\frac{1}{k^{2}} 
+ \sum_{k=s_{L}}^{\infty}\frac{1}{k^{2}}. 
\label{S1_bound}  
\end{IEEEeqnarray} 
On the other hand, for $S_{2}(\Omega)$  
\begin{equation} \label{S2_bound}
S_{2}(\Omega) 
\leq \sum_{k=\lfloor L\Omega\rfloor+1}^{L}\frac{1}{(2L-k)^{2}} 
< \sum_{k=L}^{\infty}\frac{1}{k^{2}}. 
\end{equation}

Combining (\ref{G_upper_bound2}), (\ref{sinc_bound2}), (\ref{S1_bound}), 
and (\ref{S2_bound}), from Lemma~\ref{lemma_appen0} we find that 
$s_{L}G(\Omega)$ is uniformly 
bounded for all $\Omega\in\mathcal{D}_{L}$ and $\Delta\in(0,1/2]$ 
in the limit $L, N\to\infty$ with $\Delta=L/N$ fixed.

\section{Proof of Theorem~\ref{theorem1}}
\label{proof_theorem1}
\subsection{Asymptotically Equivalent Matrices} 
In order to prove Theorem~\ref{theorem1}, we first introduce the notion 
of asymptotically equivalent matrices~\cite[Chapter~2]{Gray06}. 
After defining two norms on the space of $N\times M$ complex matrices, 
we present the definition of asymptotically equivalent matrices.  

\begin{definition}
The operator norm of $\boldsymbol{A}\in\mathbb{C}^{N\times M}$ is defined as 
\begin{equation} \label{operator_norm}
\|\boldsymbol{A}\|
= \sup_{\boldsymbol{x}\in\mathbb{C}^{M}:\boldsymbol{x}\neq\boldsymbol{0}}
\frac{\|\boldsymbol{A}\boldsymbol{x}\|}{\|\boldsymbol{x}\|}. 
\end{equation}
\end{definition} 

\begin{definition}
The normalized Frobenius norm of 
$\boldsymbol{A}\in\mathbb{C}^{N\times M}$ is defined as 
\begin{equation} 
\|\boldsymbol{A}\|_{\mathrm{F}} 
= \sqrt{N^{-1}\mathrm{Tr}(\boldsymbol{A}\boldsymbol{A}^{\mathrm{H}})}. 
\end{equation}
\end{definition}

\begin{definition}
Two matrices $\boldsymbol{A}, \boldsymbol{B}\in\mathbb{C}^{N\times M}$ 
are called asymptotically equivalent matrices if the operator norms 
$\|\boldsymbol{A}\|$ and $\|\boldsymbol{B}\|$ are uniformly bounded 
for all $N$ and $M$, and if the normalized Frobenius norm 
$\|\boldsymbol{A}-\boldsymbol{B}\|_{\mathrm{F}}$ tends to zero in the 
large matrix limit $N, M\to\infty$ with $\beta=M/N$ kept constant. 
\end{definition}

The purpose of this section is to prove the following theorem: 
\begin{theorem}[\cite{Gray06}] \label{theorem_appen1}
For two positive semi-definite Hermitian matrices 
$\boldsymbol{A}, \boldsymbol{B}\in\mathbb{C}^{N\times N}$, 
let $\{\lambda_{n}\geq0\}$ and 
$\{\tilde{\lambda}_{n}\geq0\}$ denote the eigenvalues of 
$\boldsymbol{A}$ and $\boldsymbol{B}$, respectively. 
If $\boldsymbol{A}$ and $\boldsymbol{B}$ are asymptotically 
equivalent, then 
\begin{equation} \label{target_formula}
J = \frac{1}{N}\sum_{n=0}^{N-1}\left\{
 \psi(\lambda_{n}) - \psi(\tilde{\lambda}_{n})
\right\} \to 0, 
\end{equation}
in the large matrix limit, for any continuous function $\psi$. 
\end{theorem}

In order to prove Theorem~\ref{theorem_appen1}, we first present several 
properties of the two norms. 

\begin{property} \label{property2} 
\begin{itemize}
\item The operator norm $\|\boldsymbol{A}\|$ is equal to the maximum singular 
value of $\boldsymbol{A}$. Furthermore, 
$\|\boldsymbol{A}_{1}\|\leq\|\boldsymbol{A}\|$ for any submatrix 
$\boldsymbol{A}_{1}$ of $\boldsymbol{A}$. 

\item The operator norm $\|\cdot\|$ is submultiplicative, 
\begin{equation}
\|\boldsymbol{A}\boldsymbol{B}\|\leq \|\boldsymbol{A}\|\|\boldsymbol{B}\|. 
\end{equation}

\item For $\boldsymbol{A}\in\mathbb{C}^{N\times L}$ and 
$\boldsymbol{B}\in\mathbb{C}^{L\times M}$, 
\begin{equation}
\|\boldsymbol{A}\boldsymbol{B}\|_{\mathrm{F}} 
\leq \sqrt{\frac{L}{N}}
\|\boldsymbol{A}\|\|\boldsymbol{B}\|_{\mathrm{F}}. 
\end{equation}

\item For the square case $N=M$, 
\begin{equation}
\frac{1}{N}|\mathrm{Tr}(\boldsymbol{A})| 
\leq \|\boldsymbol{A}\|_{\mathrm{F}}. 
\end{equation}
\end{itemize}
\end{property}
\begin{IEEEproof}
The former statement in the first property is well known. In proving the 
latter statement, without loss of generality, we assume 
\begin{equation}
\boldsymbol{A} 
= 
\begin{pmatrix}
\boldsymbol{A}_{1} & \boldsymbol{A}_{2} \\ 
\boldsymbol{A}_{3} & \boldsymbol{A}_{4} 
\end{pmatrix}. 
\end{equation}
Let $M_{1}$ denote the number of columns of 
the submatrix $\boldsymbol{A}_{1}$. By restricting the $m$th elements of 
$\boldsymbol{x}$ in (\ref{operator_norm}) to zero for $m\geq M_{1}$, 
we obtain the lower bound 
\begin{IEEEeqnarray}{rl}
\|\boldsymbol{A}\|
\geq& \sup_{\tilde{\boldsymbol{x}}\in\mathbb{C}^{M_{1}}:
\tilde{\boldsymbol{x}}\neq \boldsymbol{0}}
\frac{1}{\|\tilde{\boldsymbol{x}}\|}
\left\|
 \begin{pmatrix}
 \boldsymbol{A}_{1} \\ 
 \boldsymbol{A}_{3}
 \end{pmatrix}
 \tilde{\boldsymbol{x}}
\right\|
\nonumber \\ 
=& \sup_{\boldsymbol{y}\in\mathbb{C}^{N}:\boldsymbol{y}\neq \boldsymbol{0}}
\frac{\|(\boldsymbol{A}_{1}^{\mathrm{T}}, \boldsymbol{A}_{3}^{\mathrm{T}})
\boldsymbol{y}\|}{\|\boldsymbol{y}\|}. 
\end{IEEEeqnarray}
Repeating the same argument yields 
$\|\boldsymbol{A}\|\geq\|\boldsymbol{A}_{1}\|$. 

The second property is trivial for $\boldsymbol{B}=\boldsymbol{O}$. 
For non-zero matrices $\boldsymbol{B}$, it follows from 
\begin{equation}
\|\boldsymbol{A}\boldsymbol{B}\| 
= \sup_{\boldsymbol{x}\neq\boldsymbol{0}}
\frac{\|\boldsymbol{A}(\boldsymbol{B}\boldsymbol{x})\|}
{\|\boldsymbol{B}\boldsymbol{x}\|}
\frac{\|\boldsymbol{B}\boldsymbol{x}\|}{\|\boldsymbol{x}\|}
\leq \sup_{\boldsymbol{y}\neq\boldsymbol{0}}
\frac{\|\boldsymbol{A}\boldsymbol{y}\|}{\|\boldsymbol{y}\|}
\sup_{\boldsymbol{x}\neq\boldsymbol{0}}
\frac{\|\boldsymbol{B}\boldsymbol{x}\|}{\|\boldsymbol{x}\|}. 
\end{equation}

We shall show the third property. Let $\boldsymbol{b}_{m}$ denote the $m$th 
non-zero column of $\boldsymbol{B}$. From the definition of the 
normalized Frobenius norm, 
\begin{IEEEeqnarray}{rl}
\|\boldsymbol{A}\boldsymbol{B}\|_{\mathrm{F}}^{2} 
=& \frac{1}{N}\sum_{m}\frac{\|\boldsymbol{A}\boldsymbol{b}_{m}\|^{2}}
{\|\boldsymbol{b}_{m}\|^{2}}\|\boldsymbol{b}_{m}\|^{2} 
\nonumber \\ 
\leq& \frac{1}{N}
\|\boldsymbol{A}\|^{2}\sum_{m}\|\boldsymbol{b}_{m}\|^{2} 
\nonumber \\ 
=& \frac{L}{N}
\|\boldsymbol{A}\|^{2}\|\boldsymbol{B}\|_{\mathrm{F}}^{2}, 
\end{IEEEeqnarray}
where the inequality is due to the definition of the operator norm. 

Finally, we prove the last property for $N=M$. 
The trace $\mathrm{Tr}(\boldsymbol{A})$ 
can be regarded as the inner product of $\boldsymbol{A}$ and 
$\boldsymbol{I}_{N}$. We use the Cauchy-Schwarz inequality to obtain 
\begin{equation}
\frac{1}{N}|\mathrm{Tr}(\boldsymbol{A})| 
\leq \frac{1}{N}\sqrt{\mathrm{Tr}(\boldsymbol{A}\boldsymbol{A}^{\mathrm{H}})}
\sqrt{\mathrm{Tr}(\boldsymbol{I}_{N}^{2})} 
= \|\boldsymbol{A}\|_{\mathrm{F}}. 
\end{equation}
\end{IEEEproof}

In proving Theorem~\ref{theorem_appen1}, we use the following lemma 
for asymptotically equivalent matrices. 
\begin{lemma} \label{lemma_appen3} 
If $\boldsymbol{A}\in\mathbb{C}^{N\times L}$ and 
$\boldsymbol{B}\in\mathbb{C}^{N\times L}$ are asymptotically equivalent and 
if $\boldsymbol{C}\in\mathbb{C}^{L\times M}$ and 
$\boldsymbol{D}\in\mathbb{C}^{L\times M}$ are asymptotically equivalent, 
$\boldsymbol{A}\boldsymbol{C}$ and 
$\boldsymbol{B}\boldsymbol{D}$ are also asymptotically equivalent. 
\end{lemma}
\begin{IEEEproof}
The boundedness of the operator norms 
$\|\boldsymbol{A}\boldsymbol{C}\|$ and 
$\|\boldsymbol{B}\boldsymbol{D}\|$ follows from the second property 
in Property~\ref{property2}. 

For the normalized Frobenius norm, we use the triangle inequality to 
obtain 
\begin{IEEEeqnarray}{rl}
&\|\boldsymbol{A}\boldsymbol{C} 
- \boldsymbol{B}\boldsymbol{D}\|_{\mathrm{F}}  
\nonumber \\ 
\leq& \|\boldsymbol{A}(\boldsymbol{C}-\boldsymbol{D})\|_{\mathrm{F}} 
+ \|(\boldsymbol{A}-\boldsymbol{B})\boldsymbol{D}\|_{\mathrm{F}} 
\nonumber \\ 
\leq& 
\sqrt{\frac{L}{N}} 
\|\boldsymbol{A}\|\|\boldsymbol{C}-\boldsymbol{D}\|_{\mathrm{F}} 
+ \sqrt{\frac{L}{M}}\|\boldsymbol{D}\|
\|(\boldsymbol{A}-\boldsymbol{B})\|_{\mathrm{F}}, 
\label{lemma_appen3_bound} 
\end{IEEEeqnarray}
where the last inequality is due to the third property in 
Property~\ref{property2}. The bound~(\ref{lemma_appen3_bound}) implies 
$\|\boldsymbol{A}\boldsymbol{C} 
- \boldsymbol{B}\boldsymbol{D}\|_{\mathrm{F}}\to0$.   
\end{IEEEproof}

We are ready to prove Theorem~\ref{theorem_appen1}. 
\begin{IEEEproof}[Proof of Theorem~\ref{theorem_appen1}]
We first show Theorem~\ref{theorem_appen1} for the case $\psi(x)=x^{k}$ with 
non-negative integers $k$. Since $\{\lambda_{n}^{k}\}$ and 
$\{\tilde{\lambda}_{n}^{k}\}$ are the eigenvalues of 
$\boldsymbol{A}^{k}$ and $\boldsymbol{B}^{k}$, 
(\ref{target_formula}) reduces to 
\begin{equation} 
|J| = \frac{1}{N}\left|
 \mathrm{Tr}\left(
  \boldsymbol{A}^{k} 
  - \boldsymbol{B}^{k}
 \right)
\right|
\leq \left\|  
 \boldsymbol{A}^{k} - \boldsymbol{B}^{k}
\right\|_{\mathrm{F}},  \label{target_formula_power} 
\end{equation} 
where we have used the last property in Property~\ref{property2}. 
Using Lemma~\ref{lemma_appen3} repeatedly, we find that 
$\boldsymbol{A}^{k}$ and $\boldsymbol{B}^{k}$ are asymptotically 
equivalent. Thus, (\ref{target_formula_power}) tends to zero. 

It is straightforward to confirm that Theorem~\ref{theorem_appen1} is correct 
for any polynomial $\psi(x)$. For the general case, we use the Weierstrass 
approximation theorem~\cite{Gray06}. 
Since $\{\lambda_{n}\}$ and $\{\tilde{\lambda}_{n}\}$ 
are bounded, the domain of the continuous function $\psi$ can be restricted 
to an interval $[0, \lambda_{\mathrm{max}}]$. Thus, for any $\epsilon>0$ there 
exists some polynomial $p(x)$ such that 
\begin{equation} \label{approximation_theorem}
\sup_{x\in[0, \lambda_{\mathrm{max}}]}|\psi(x)-p(x)|<\frac{\epsilon}{3}. 
\end{equation}
For these $\epsilon$ and $p(x)$, we have proved 
that there are some $N_{0}\in\mathbb{N}$ such that 
\begin{equation} \label{polynomial_theorem}
\left|
 \frac{1}{N}\sum_{n=0}^{N-1}\left\{
  p(\lambda_{n}) - p(\tilde{\lambda}_{n})
\right\}
\right| < \frac{\epsilon}{3}, 
\end{equation} 
for all $N>N_{0}$. We use the triangle inequality for 
(\ref{target_formula}) to obtain
\begin{IEEEeqnarray}{rl}
|J|<& \frac{1}{N}\sum_{n=0}^{N-1}\left|
 \psi(\lambda_{n}) - p(\lambda_{n})
\right| 
+ \frac{1}{N}\sum_{n=0}^{N-1}\left|
 p(\tilde{\lambda}_{n}) - \psi(\tilde{\lambda}_{n})
\right|
\nonumber \\ 
& + \left|
 \frac{1}{N}\sum_{n=0}^{N-1}\left\{
  p(\lambda_{n}) - p(\tilde{\lambda}_{n})
\right\}
\right|
<\epsilon, 
\end{IEEEeqnarray}
where we have used (\ref{approximation_theorem}) and 
(\ref{polynomial_theorem}). 
\end{IEEEproof}

\subsection{Application of Theorem~\ref{theorem_appen1}} 
For notational convenience, the subscripts of 
$\boldsymbol{\Sigma}_{M}$, 
$\boldsymbol{G}_{\Delta_{\mathrm{t}}, \Delta_{\mathrm{r}}}$, and of  
$\tilde{\boldsymbol{G}}_{\Delta_{\mathrm{t}}, \Delta_{\mathrm{r}}}$ are omitted.  
We use Theorem~\ref{theorem_appen1} to prove Theorem~\ref{theorem1}. 
Let $\{\lambda_{n}\geq0\}$ and $\{\tilde{\lambda}_{n}\geq0\}$ denote the 
eigenvalues of the two positive semi-definite Hermitian matrices 
$\boldsymbol{A}= \boldsymbol{G}\boldsymbol{\Sigma}\boldsymbol{G}^{\mathrm{H}}$ 
and $\boldsymbol{B}=\tilde{\boldsymbol{G}}\boldsymbol{\Sigma}
\tilde{\boldsymbol{G}}^{\mathrm{H}}$, respectively. 
the left-hand side (LHS) of (\ref{asym_capacity}) is equivalent to 
\begin{equation} \label{LHS}
\frac{1}{2\Delta_{\mathrm{r}}\mathrm{min}\{1, \alpha\}}\left|
 \frac{1}{N}\sum_{n=0}^{N-1}\left\{
  \log(1+\tilde{\gamma}\lambda_{n}) 
  - \log(1+\tilde{\gamma}\tilde{\lambda}_{n})
 \right\}
\right|. 
\end{equation}

From Theorem~\ref{theorem_appen1}, the quantity (\ref{LHS}) tends to zero  
if $\boldsymbol{A}$ and $\boldsymbol{B}$ 
are asymptotically equivalent for all $\Delta_{\mathrm{t}}, 
\Delta_{\mathrm{r}}\in(0,1/2]$, covariance matrices 
$\boldsymbol{\Sigma}\in\mathfrak{S}_{M}$, and all channel instances 
$\mathcal{C}\in\mathfrak{C}$ in the large-system limit. 

From Assumption~\ref{assumption2} and Lemma~\ref{lemma_appen3}, 
it is sufficient to prove that 
$(2L_{\mathrm{t}})^{-1/2}\boldsymbol{G}$ and 
$(2L_{\mathrm{t}})^{-1/2}\tilde{\boldsymbol{G}}$ are asymptotically equivalent. 
Assumption~\ref{assumption1} implies the uniform boundedness of 
$\|(2L_{\mathrm{t}})^{-1/2}\boldsymbol{G}\|$ for fixed $\alpha$. Furthermore, 
$\|(2L_{\mathrm{t}})^{-1/2}\tilde{\boldsymbol{G}}\|$ is also uniformly bounded 
from the first property in Property~\ref{property2}. 

Next, we shall upper-bound the normalized Frobenius norm 
$\|(2L_{\mathrm{t}})^{-1/2}(\boldsymbol{G}
-\tilde{\boldsymbol{G}})\|_{\mathrm{F}}$. 
From the definition~(\ref{hat_G}) of $\tilde{\boldsymbol{G}}$, we obtain 
\begin{equation}
\|(2L_{\mathrm{t}})^{-1/2}(\boldsymbol{G}-
\tilde{\boldsymbol{G}})\|_{\mathrm{F}}^{2}  
= J_{1} + J_{2} + J_{3}, 
\end{equation}
with 
\begin{equation}
J_{1} 
= \frac{\Delta_{\mathrm{r}}}{2L_{\mathrm{t}}L_{\mathrm{r}}}
\sum_{n\in\mathcal{N}}\sum_{m\notin\mathcal{M}}|g_{n,m}|^{2},  
\end{equation}
\begin{equation}
J_{2} 
= \frac{\Delta_{\mathrm{r}}}{2L_{\mathrm{t}}L_{\mathrm{r}}}
\sum_{n\notin\mathcal{N}}\sum_{m\in\mathcal{M}}|g_{n,m}|^{2}, 
\end{equation}
\begin{equation}
J_{3} 
= \frac{\Delta_{\mathrm{r}}}{2L_{\mathrm{t}}L_{\mathrm{r}}}
\sum_{n\notin\mathcal{N}}\sum_{m\notin\mathcal{M}}|g_{n,m}|^{2}. 
\end{equation}

We first evaluate $J_{1}$. Using (\ref{channel_gain}) yields 
\begin{IEEEeqnarray}{rl}
J_{1} 
< \frac{\Delta_{\mathrm{r}}}{2L_{\mathrm{t}}L_{\mathrm{r}}}
\sum_{n=0}^{N-1}&\sum_{m\notin\mathcal{M}}|g_{n,m}|^{2} 
\nonumber \\ 
= 2\Delta_{\mathrm{r}}\int a(p)&a^{*}(p')
I_{1}(\Omega_{\mathrm{r}}(p),\Omega_{\mathrm{r}}(p');
L_{\mathrm{r}},\Delta_{\mathrm{r}}) 
\nonumber \\
&\cdot I_{2}(\Omega_{\mathrm{t}}(p),\Omega_{\mathrm{t}}(p');
L_{\mathrm{t}},\Delta_{\mathrm{t}})dpdp', 
\end{IEEEeqnarray}
with (\ref{sum}) and (\ref{partial_sum}). 
From Lemmas~\ref{lemma1} and \ref{lemma2}, $s_{L}J_{1}$ is uniformly bounded 
for all $\Delta_{\mathrm{t}}, \Delta_{\mathrm{r}}\in(0,1/2]$ and 
all channel instances $\mathcal{C}\in\mathfrak{C}$ 
in the large-system limit.  

Repeating the same argument for $J_{2}$, we find that 
$s_{L}J_{2}$ is uniformly bounded. Similarly, it is possible to prove that 
$s_{L}^{2}J_{3}$ is uniformly bounded in the large-system limit. 
Combining these observations, we arrive at 
$\|(2L_{\mathrm{t}})^{-1/2}
(\boldsymbol{G}-\tilde{\boldsymbol{G}})\|_{\mathrm{F}}\to0$ 
uniformly for all $\Delta_{\mathrm{t}}, \Delta_{\mathrm{r}}\in(0,1/2]$ and 
all channel instances $\mathcal{C}\in\mathfrak{C}$ 
in the large-system limit.   

\section{Proof of Theorem~\ref{theorem2}} 
\label{proof_theorem2} 
Theorem~\ref{theorem2} is proved by repeating the proof of 
Theorem~\ref{theorem1}, while Lemma~\ref{lemma3} is used instead of 
Lemma~\ref{lemma2}. Thus, we use the same notation as in the proof of 
Theorem~\ref{theorem1}. 

For notational convenience, define the shrunk channel matrix 
$\bar{\boldsymbol{G}}_{\Delta_{\mathrm{t}},\Delta_{\mathrm{r}}}
\in\mathbb{C}^{(2L_{\mathrm{r}}+1)\times(2L_{\mathrm{t}}+1)}$ obtained by eliminating 
the all-zero columns and rows from the channel matrix~(\ref{hat_G}), and 
subsequently by moving the $L_{\mathrm{t}}$th column and $L_{\mathrm{r}}$th 
row to the last positions. Thus, we should consider the covariance matrix 
$\bar{\boldsymbol{\Sigma}}_{2L_{\mathrm{t}}+1}$ obtained by moving the 
$L_{\mathrm{t}}$th column and row of $\boldsymbol{\Sigma}_{2L_{\mathrm{t}}+1}$ 
to the last positions, as well as the extended matrix 
$\bar{\boldsymbol{G}}_{1/2,1/2}\in\mathbb{C}^{(2L_{\mathrm{r}}+1)\times(2L_{\mathrm{t}}+1)}$ 
obtained by inserting all-zero vectors after the last column and row of 
the matrix $\boldsymbol{G}_{1/2,1/2}$. 
 
Let $\{\lambda_{n}\geq0\}$ and $\{\tilde{\lambda}_{n}\geq0\}$ denote the 
eigenvalues of the two Hermitian matrices 
$\boldsymbol{A}=\bar{\boldsymbol{G}}_{1/2,1/2}
\boldsymbol{E}_{2L_{\mathrm{t}},1}(\boldsymbol{\Sigma}_{2L_{\mathrm{t}}})
\bar{\boldsymbol{G}}_{1/2,1/2}^{\mathrm{H}}$ and 
$\boldsymbol{B}
=\bar{\boldsymbol{G}}_{\Delta_{\mathrm{t}},\Delta_{\mathrm{r}}}
\bar{\boldsymbol{\Sigma}}_{2L_{\mathrm{t}}+1}
\bar{\boldsymbol{G}}_{\Delta_{\mathrm{t}},\Delta_{\mathrm{r}}}^{\mathrm{H}}$, 
respectively. The LHS of (\ref{capacity_difference}) is equivalent to 
\begin{equation}
\frac{1}{\min\{1, \alpha\}}
\left|
 \frac{1}{2L_{\mathrm{r}}}\sum_{n=0}^{2L_{\mathrm{r}}}\left\{
  \log(1+\tilde{\gamma}\lambda_{n}) 
  - \log(1+\tilde{\gamma}\tilde{\lambda}_{n})
 \right\}
\right|. 
\end{equation}

Assumption~\ref{assumption2} and the condition~(\ref{condition}) imply that 
$(2L_{\mathrm{t}}+1)\boldsymbol{E}_{2L_{\mathrm{t}},1}
(\boldsymbol{\Sigma}_{2L_{\mathrm{t}}})$ and 
$(2L_{\mathrm{t}}+1)\bar{\boldsymbol{\Sigma}}_{2L_{\mathrm{t}}+1}$ are asymptotically 
equivalent.  
From the proof of Theorem~\ref{theorem1}, it is sufficient to prove that 
$K=\|(2L_{\mathrm{t}}+1)^{-1/2}(\bar{\boldsymbol{G}}_{1/2,1/2}
- \bar{\boldsymbol{G}}_{\Delta_{\mathrm{t}},\Delta_{\mathrm{r}}})\|_{\mathrm{F}}^{2}$ 
uniformly converges to zero in the large-system limit. 
From (\ref{channel_gain}) and the definition of 
$\bar{\boldsymbol{G}}_{\Delta_{\mathrm{t}},\Delta_{\mathrm{r}}}
\in\mathbb{C}^{(2L_{\mathrm{r}}+1)\times (2L_{\mathrm{t}}+1)}$, 
\begin{IEEEeqnarray}{rl} 
K &= \frac{4L_{\mathrm{t}}L_{\mathrm{r}}}{(2L_{\mathrm{t}}+1)(2L_{\mathrm{r}}+1)} 
\nonumber \\ 
\cdot&\int a(p)a^{*}(p')\sum_{n=0}^{2L_{\mathrm{r}}}
\sum_{m=0}^{2L_{\mathrm{t}}}\tilde{D}_{n,m}(p)
\tilde{D}_{n,m}^{*}(p')dpdp', 
\label{J4}
\end{IEEEeqnarray}
with 
\begin{IEEEeqnarray}{rl}
\tilde{D}_{n,m}(p)
=& (1-\delta_{n,2L_{\mathrm{r}}})(1-\delta_{m,2L_{\mathrm{t}}})f_{1/2}(n,p)f_{1/2}(m,p) 
\nonumber \\ 
&- f_{\Delta_{\mathrm{r}}}(\kappa_{L_{\mathrm{r}},\Delta_{\mathrm{r}}}(n),p)
f_{\Delta_{\mathrm{t}}}(\kappa_{L_{\mathrm{t}},\Delta_{\mathrm{t}}}(m),p), 
\label{tilde_D}
\end{IEEEeqnarray}
with the mapping~$\kappa_{L,\Delta}(k)$ given by (\ref{kappa}). 
In (\ref{tilde_D}), the abbreviations $f_{\Delta_{\mathrm{r}}}(n,p)=
f_{L_{\mathrm{r}},\Delta_{\mathrm{r}}}(n/L_{\mathrm{r}} - \Omega_{\mathrm{r}}(p))$
and $f_{\Delta_{\mathrm{t}}}(m,p)=
f_{L_{\mathrm{t}},\Delta_{\mathrm{t}}}(m/L_{\mathrm{t}} - \Omega_{\mathrm{t}}(p))$ have 
been introduced. Note that $n$ and $m$ are not dummy variables but the 
indices of receive and transmit antennas, respectively.  

Substituting the identity 
\begin{IEEEeqnarray}{rl}
\tilde{D}_{n,m}(p)
=& [(1-\delta_{n,2L_{\mathrm{r}}})f_{1/2}(n,p) 
- f_{\Delta_{\mathrm{r}}}(\kappa_{L_{\mathrm{r}},\Delta_{\mathrm{r}}}(n),p)]
\nonumber \\ 
&\cdot(1-\delta_{m,2L_{\mathrm{t}}})f_{1/2}(m,p) 
\nonumber \\ 
&+ [(1-\delta_{m,2L_{\mathrm{t}}})f_{1/2}(m,p) 
- f_{\Delta_{\mathrm{t}}}(\kappa_{L_{\mathrm{t}},\Delta_{\mathrm{t}}}(m),p)]
\nonumber \\
&\cdot f_{\Delta_{\mathrm{r}}}(\kappa_{L_{\mathrm{r}},\Delta_{\mathrm{r}}}(n),p)
\end{IEEEeqnarray}
into (\ref{J4}), we have 
\begin{equation}
K = \frac{4L_{\mathrm{t}}L_{\mathrm{r}}}{(2L_{\mathrm{t}}+1)(2L_{\mathrm{r}}+1)} 
\left(
 K_{1} + K_{2} + K_{2}^{*} + K_{3}
\right),
\end{equation}
with 
\begin{IEEEeqnarray}{r}
K_{1} 
= \int a(p)a^{*}(p')
I_{3}(\Omega_{\mathrm{r}}(p),\Omega_{\mathrm{r}}(p');L_{\mathrm{r}},
\Delta_{\mathrm{r}})
\nonumber \\ 
\cdot 
I_{1}(\Omega_{\mathrm{t}}(p),\Omega_{\mathrm{t}}(p');L_{\mathrm{t}},1/2)dpdp', 
\end{IEEEeqnarray}
\begin{IEEEeqnarray}{rl}
K_{2} 
= \int& a(p)a^{*}(p')
\sum_{n=0}^{2L_{\mathrm{r}}}
f_{\Delta_{\mathrm{r}}}^{*}(\kappa_{L_{\mathrm{r}},\Delta_{\mathrm{r}}}(n),p') 
\nonumber \\ 
&\cdot[(1-\delta_{n,2L_{\mathrm{r}}})f_{1/2}(n,p) 
- f_{\Delta_{\mathrm{r}}}(\kappa_{L_{\mathrm{r}},\Delta_{\mathrm{r}}}(n),p)]
\nonumber \\ 
&\cdot\sum_{m=0}^{2L_{\mathrm{t}}-1}f_{1/2}(m,p) 
\nonumber \\
&\cdot[f_{1/2}(m,p') 
- f_{\Delta_{\mathrm{t}}}(\kappa_{L_{\mathrm{t}},\Delta_{\mathrm{t}}}(m),p')]^{*}dpdp', 
\end{IEEEeqnarray}
\begin{IEEEeqnarray}{rl}
K_{3} 
&= \int a(p)a^{*}(p')
I_{3}(\Omega_{\mathrm{t}}(p),\Omega_{\mathrm{t}}(p');L_{\mathrm{t}},
\Delta_{\mathrm{t}})
\nonumber \\ 
\cdot&\sum_{n=0}^{2L_{\mathrm{r}}}
f_{\Delta_{\mathrm{r}}}(\kappa_{L_{\mathrm{r}},\Delta_{\mathrm{r}}}(n),p)
f_{\Delta_{\mathrm{r}}}^{*}(\kappa_{L_{\mathrm{r}},\Delta_{\mathrm{r}}}(n),p')dpdp', 
\end{IEEEeqnarray}
where $I_{1}$ and $I_{3}$ are given by (\ref{sum}) and 
(\ref{difference_sum}), respectively. 

We first upper-bound $K_{1}$ and $K_{3}$. Under Assumption~\ref{assumption1}, 
Lemmas~\ref{lemma1} and \ref{lemma3} imply that $s_{L}|K_{1}|$ is uniformly 
bounded for all channel instances and antenna separations in the large-system  
limit. Similarly, we find the uniform boundedness of $s_{L}|K_{3}|$ 
from the upper bound 
\begin{IEEEeqnarray}{rl}
&\left|
 \sum_{n=0}^{2L_{\mathrm{r}}}
 f_{\Delta_{\mathrm{r}}}(\kappa_{L_{\mathrm{r}},\Delta_{\mathrm{r}}}(n),p)
 f_{\Delta_{\mathrm{r}}}(\kappa_{L_{\mathrm{r}},\Delta_{\mathrm{r}}}(n),p')^{*} 
\right|^{2}
\nonumber \\ 
\leq& 
\sum_{n=0}^{2L_{\mathrm{r}}}
|f_{\Delta_{\mathrm{r}}}(\kappa_{L_{\mathrm{r}},\Delta_{\mathrm{r}}}(n),p)|^{2}
\sum_{n'=0}^{2L_{\mathrm{r}}}
|f_{\Delta_{\mathrm{r}}}(\kappa_{L_{\mathrm{r}},\Delta_{\mathrm{r}}}(n'),p')|^{2} 
\nonumber \\
<& |I_{1}(\Omega_{\mathrm{r}}(p),\Omega_{\mathrm{r}}(p);
L_{\mathrm{r}},\Delta_{\mathrm{r}})|^{2}. 
\end{IEEEeqnarray}
In the derivation of the first inequality, we have used the 
Cauchy-Schwarz inequality. 
 
We next evaluate $|K_{2}|$. Using the Cauchy-Schwarz inequality, we have 
\begin{IEEEeqnarray}{rl}
|K_{2}| 
\leq& \int|a(p)||a(p')|\left\{
 \sum_{n=0}^{2L_{\mathrm{r}}}
 |f_{\Delta_{\mathrm{r}}}(\kappa_{L_{\mathrm{r}},\Delta_{\mathrm{r}}}(n),p')|^{2}
\right.
\nonumber \\ 
&\cdot I_{3}(\Omega_{\mathrm{r}}(p),\Omega_{\mathrm{r}}(p);L_{\mathrm{r}},
 \Delta_{\mathrm{r}})
|I_{1}(\Omega_{\mathrm{t}}(p),\Omega_{\mathrm{t}}(p);L_{\mathrm{t}},1/2)|
\nonumber \\
&\cdot
 I_{3}(\Omega_{\mathrm{t}}(p'),\Omega_{\mathrm{t}}(p');L_{\mathrm{t}},
 \Delta_{\mathrm{t}})
\biggr\}^{1/2}dpdp'. \label{K2_bound} 
\end{IEEEeqnarray}
Upper-bounding the sum in (\ref{K2_bound}) yields  
\begin{IEEEeqnarray}{rl}
|K_{2}| 
<& \int|a(p)|\{
|I_{3}(\Omega_{\mathrm{r}}(p),\Omega_{\mathrm{r}}(p);L_{\mathrm{r}},
\Delta_{\mathrm{r}})|
\nonumber \\ 
&\cdot 
|I_{1}(\Omega_{\mathrm{t}}(p),\Omega_{\mathrm{t}}(p);L_{\mathrm{t}},1/2)|
\}^{1/2}dp 
\nonumber \\ 
\cdot&\int|a(p')|\{
|I_{1}(\Omega_{\mathrm{r}}(p'),\Omega_{\mathrm{r}}(p');L_{\mathrm{r}},
\Delta_{\mathrm{r}})|
\nonumber \\ 
&\cdot 
|I_{3}(\Omega_{\mathrm{t}}(p'),\Omega_{\mathrm{t}}(p');L_{\mathrm{t}},
\Delta_{\mathrm{t}})|
\}^{1/2}dp'. 
\end{IEEEeqnarray}
Thus, $s_{L}|K_{2}|$ is also uniformly bounded. Combining these observations, 
we arrive at Theorem~\ref{theorem2}. 


\ifCLASSOPTIONcaptionsoff
  \newpage
\fi



\bibliographystyle{IEEEtran}
\bibliography{IEEEabrv,kt-it2016_1}

\begin{thebibliography}{10}
\providecommand{\url}[1]{#1}
\csname url@samestyle\endcsname
\providecommand{\newblock}{\relax}
\providecommand{\bibinfo}[2]{#2}
\providecommand{\BIBentrySTDinterwordspacing}{\spaceskip=0pt\relax}
\providecommand{\BIBentryALTinterwordstretchfactor}{4}
\providecommand{\BIBentryALTinterwordspacing}{\spaceskip=\fontdimen2\font plus
\BIBentryALTinterwordstretchfactor\fontdimen3\font minus
  \fontdimen4\font\relax}
\providecommand{\BIBforeignlanguage}[2]{{%
\expandafter\ifx\csname l@#1\endcsname\relax
\typeout{** WARNING: IEEEtran.bst: No hyphenation pattern has been}%
\typeout{** loaded for the language `#1'. Using the pattern for}%
\typeout{** the default language instead.}%
\else
\language=\csname l@#1\endcsname
\fi
#2}}
\providecommand{\BIBdecl}{\relax}
\BIBdecl

\bibitem{Takeuchi16}
K.~Takeuchi, ``Asymptotic optimality of massive {MIMO} systems using densely
  spaced transmit antennas,'' \emph{{\em submitted to} ISIT 2016}, 2016.

\bibitem{Marzetta06}
T.~L. Marzetta, ``How much training is required for multiuser {MIMO}?'' in
  \emph{Proc. 40th Asilomar Conf. on Signals, Systems and Computers}, Pacific
  Grove, CA, Oct.--Nov. 2006, pp. 359--363.

\bibitem{Marzetta10}
------, ``Noncooperative cellular wireless with unlimited numbers of base
  station antennas,'' \emph{IEEE Trans. Wireless Commun.}, vol.~9, no.~11, pp.
  3590--3600, Nov. 2010.

\bibitem{Rusek13}
F.~Rusek, D.~Persson, B.~K. Lau, E.~G. Larsson, T.~L. Marzetta, O.~Edfors, and
  F.~Tufvesson, ``Scaling up {MIMO}: Opportunities and challenges with very
  large arrays,'' \emph{{IEEE} Signal Process. Mag.}, vol.~30, no.~1, pp.
  40--60, Jan. 2013.

\bibitem{Larsson14}
E.~G. Larsson, O.~Edfors, F.~Tufvesson, and T.~L. Marzetta, ``Massive {NIMO}
  for next generation wireless systems,'' \emph{{IEEE} Commun. Mag.}, vol.~52,
  no.~2, pp. 186--195, Feb. 2014.

\bibitem{Telatar99}
E.~Telatar, ``Capacity of multi-antenna {Gaussian} channels,'' \emph{Euro.
  Trans. Telecommun.}, vol.~10, no.~6, pp. 585--595, Nov.--Dec. 1999.

\bibitem{Wallace03}
J.~W. Wallace, M.~A. Jensen, A.~L. Swindlehurst, and B.~D. Jeffs,
  ``Experimental characterization of the {MIMO} wireless channel: Data
  acquisition and analysis,'' \emph{{IEEE} Trans. Wireless Commun.}, vol.~2,
  no.~2, pp. 335--343, Mar. 2003.

\bibitem{Chizhik03}
D.~Chizhik, J.~Ling, P.~W. Wolniansky, R.~A. Valenzuela, N.costa, and K.~Huber,
  ``Multiple-input-multiple-output measurements and modeling in {Manhattan},''
  \emph{{IEEE} J. Sel. Areas Commun.}, vol.~21, no.~3, pp. 321--331, Apr. 2003.

\bibitem{Jensen04}
M.~A. Jensen and J.~W. Wallace, ``A review of antennas and propagation for
  {MIMO} wireless communications,'' \emph{{IEEE} Trans. Antennas Propag.},
  vol.~52, no.~11, pp. 2810--2824, Nov. 2004.

\bibitem{Janaswamy02}
R.~Janaswamy, ``Effect of element mutual coupling on the capacity of fixed
  length linear arrays,'' \emph{{IEEE} Antennas Wireless Propag. Lett.},
  vol.~1, no.~1, pp. 157--160, 2002.

\bibitem{Wallace04}
J.~W. Wallace and M.~A. Jensen, ``Mutual coupling in {MIMO} wireless systems: A
  rigorous network theory analysis,'' \emph{{IEEE} Trans. Wireless Commun.},
  vol.~3, no.~4, pp. 1317--1325, Jul. 2004.

\bibitem{Lau06}
B.~K. Lau, J.~B. Andersen, G.~Kristensson, and A.~F. Molisch, ``Impact of
  matching network on bandwidth of compact antenna arrays,'' \emph{{IEEE}
  Trans. Antennas Propag.}, vol.~54, no.~11, pp. 3225--3238, Nov. 2006.

\bibitem{Volmer08}
C.~Volmer, J.~Weber, R.~Stephan, K.~Blau, and M.~A. Hein, ``An eigen-analysis
  of compact antenna arrays and its application to port decoupling,''
  \emph{{IEEE} Trans. Antennas Propag.}, vol.~56, no.~2, pp. 360--370, Feb.
  2008.

\bibitem{Poon05}
A.~S.~Y. Poon, R.~W. Brodersen, and D.~N.~C. Tse, ``Degrees of freedom in
  multiple-antenna channels: A signal space approach,'' \emph{{IEEE} Trans.
  Inf. Theory}, vol.~51, no.~2, pp. 523--536, Feb. 2005.

\bibitem{Poon06}
A.~S.~Y. Poon, D.~N.~C. Tse, and R.~W. Brodersen, ``Impact of scattering on the
  capacity, diversity, and propagation range of multiple-antenna channels,''
  \emph{{IEEE} Trans. Inf. Theory}, vol.~52, no.~3, pp. 1087--1100, Mar. 2006.

\bibitem{Mazo75}
J.~E. Mazo, ``Faster-than-{Nyquist} signaling,'' \emph{Bell Syst. Tech. J.},
  vol.~54, no.~8, pp. 1451--1462, Oct. 1975.

\bibitem{Rusek09}
F.~Rusek and J.~B. Anderson, ``Constrained capacities for faster-than-{Nyquist}
  signaling,'' \emph{{IEEE} Trans. Inf. Theory}, vol.~55, no.~2, pp. 764--775,
  Feb. 2009.

\bibitem{Yoo10}
Y.~G. Yoo and J.~H. Cho, ``Asymptotic optimality of binary
  faster-than-{Nyquist} signaling,'' \emph{{IEEE} Commun. Lett.}, vol.~14,
  no.~9, pp. 788--790, Sep. 2010.

\bibitem{Cover06}
T.~M. Cover and J.~A. Thomas, \emph{Elements of Information Theory},
  2nd~ed.\hskip 1em plus 0.5em minus 0.4em\relax New Jersey: Wiley, 2006.

\bibitem{Verdu99}
S.~Verd\'u and S.~{Shamai (Shitz)}, ``Spectral efficiency of {CDMA} with random
  spreading,'' \emph{{IEEE} Trans. Inf. Theory}, vol.~45, no.~2, pp. 622--640,
  Mar. 1999.

\bibitem{Tse99}
D.~N.~C. Tse and S.~V. Hanly, ``Linear multiuser receivers: effective
  interference, effective bandwidth and user capacity,'' \emph{{IEEE} Trans.
  Inf. Theory}, vol.~45, no.~2, pp. 641--657, Mar. 1999.

\bibitem{Tulino05}
A.~M. Tulino, A.~Lozano, and S.~Verd\'{u}, ``Impact of antenna correlation on
  the capacity of multiantenna channels,'' \emph{{IEEE} Trans. Inf. Theory},
  vol.~51, no.~7, pp. 2491--2509, Jul. 2005.

\bibitem{Dumont10}
J.~Dumont, W.~Hachem, S.~Lasaulce, P.~Loubaton, and J.~Najim, ``On the capacity
  achieving covariance matrix for {Rician} {MIMO} channels: An asymptotic
  approach,'' \emph{{IEEE} Trans. Inf. Theory}, vol.~56, no.~3, pp. 1048--1069,
  Mar. 2010.

\bibitem{Couillet11}
R.~Couillet, M.~Debbah, and J.~W. Silverstein, ``A deterministic equivalent for
  the analysis of correlated {MIMO} multiple access channels,'' \emph{{IEEE}
  Trans. Inf. Theory}, vol.~57, no.~6, pp. 3493--3514, Jun. 2011.

\bibitem{Mueller14}
R.~R. M\"uller, L.~Cottatellucci, and M.~Vehkaper\"a, ``Blind pilot
  decontamination,'' \emph{IEEE J. Sel. Topics Signal Process.}, vol.~8, no.~5,
  pp. 773--786, Oct. 2014.

\bibitem{Tanaka02}
T.~Tanaka, ``A statistical-mechanics approach to large-system analysis of
  {CDMA} multiuser detectors,'' \emph{{IEEE} Trans. Inf. Theory}, vol.~48,
  no.~11, pp. 2888--2910, Nov. 2002.

\bibitem{Moustakas03}
A.~L. Moustakas, S.~H. Simon, and A.~M. Sengupta, ``{MIMO} capacity through
  correlated channels in the presence of correlated interferers and noise: A
  (not so) large ${N}$ analysis,'' \emph{{IEEE} Trans. Inf. Theory}, vol.~49,
  no.~10, pp. 2545--2561, Oct. 2003.

\bibitem{Guo05}
D.~Guo and S.~Verd\'u, ``Randomly spread {CDMA}: Asymptotics via statistical
  physics,'' \emph{{IEEE} Trans. Inf. Theory}, vol.~51, no.~6, pp. 1983--2010,
  Jun. 2005.

\bibitem{Wen06}
C.-K. Wen, P.~Ting, and J.-T. Chen, ``Asymptotic analysis of {MIMO} wireless
  systems with spatial correlation at the receiver,'' \emph{{IEEE} Trans.
  Commun.}, vol.~54, no.~2, pp. 349--363, Feb. 2006.

\bibitem{Zaidel12}
B.~M. Zaidel, R.~R. M\"uller, A.~L. Moustakas, and R.~de~Miguel, ``Vector
  precoding for {Gaussian} {MIMO} broadcast channels: Impact of replica
  symmetry breaking,'' \emph{{IEEE} Trans. Inf. Theory}, vol.~58, no.~3, pp.
  1413--1440, Mar. 2012.

\bibitem{Takeuchi12}
K.~Takeuchi, M.~Vehkaper\"a, T.~Tanaka, and R.~R. M\"uller, ``Large-system
  analysis of joint channel and data estimation for {MIMO DS-CDMA} systems,''
  \emph{{IEEE} Trans. Inf. Theory}, vol.~58, no.~3, pp. 1385--1412, Mar. 2012.

\bibitem{Takeuchi13}
K.~Takeuchi, R.~R. M\"uller, M.~Vehkaper\"a, and T.~Tanaka, ``On an achievable
  rate of large {Rayleigh} block-fading {MIMO} channels with no {CSI},''
  \emph{{IEEE} Trans. Inf. Theory}, vol.~59, no.~10, pp. 6517--6541, Oct. 2013.

\bibitem{Girnyk14}
M.~A. Girnyk, M.~Vehkaper\"a, and L.~K. Rasmussen, ``Large-system analysis of
  correlated {MIMO} multiple access channels with arbitrary signaling in the
  presence of interference,'' \emph{{IEEE} Trans. Wireless Commun.}, vol.~13,
  no.~4, pp. 2060--2073, Apr. 2014.

\bibitem{Biglieri02}
E.~Biglieri, G.~Taricco, and A.~Tulino, ``How far aways is infinity? using
  asymptotic analyses in multiple-antenna systems,'' in \emph{Proc. IEEE 7th
  Int. Symp. Spread-Spectrum Tech. {\rm \&} Appl.}, vol.~1, Prague, Czech
  Republic, Sep. 2002, pp. 1--6.

\bibitem{Mesleh08}
R.~Y. Mesleh, H.~Haas, S.~Sinanovi\'c, C.~W. Ahn, and S.~Yun, ``Spatial
  modulation,'' \emph{{IEEE} Trans. Veh. Technol.}, vol.~57, no.~4, pp.
  2228--2241, Jul. 2008.

\bibitem{Jeganathan09}
J.~Jeganathan, A.~Ghrayeb, L.~Szczecinski, and A.~Ceron, ``Space shift keying
  modulation for {MIMO} channels,'' \emph{{IEEE} Trans. Wireless Commun.},
  vol.~8, no.~7, pp. 3692--3703, Jul. 2009.

\bibitem{Renzo14}
M.~D. Renzo, H.~Haas, A.~Ghrayeb, S.~Sugiura, and L.~Hanzo, ``Spatial
  modulation for generalized {MIMO}: Challenges, oppotunities, and
  implementation,'' \emph{Proc. IEEE}, vol. 102, no.~1, pp. 56--103, Jan. 2014.

\bibitem{Takeuchi15}
K.~Takeuchi, ``Second-order optimality of generalized spatial modulation for
  {MIMO} channels with no {CSI},'' \emph{IEEE Wireless Commun. Lett.}, vol.~4,
  no.~6, pp. 613--616, Dec. 2015.

\bibitem{Visotsky01}
E.~Visotsky and U.~Madhow, ``Space-time transmit precoding with imperfect
  feedback,'' \emph{{IEEE} Trans. Inf. Theory}, vol.~47, no.~6, pp. 2632--2639,
  Sep. 2001.

\bibitem{Jafar04}
S.~A. Jafar and A.~Goldsmith, ``Transmitter optimization and optimality of
  beamforming for multiple antenna systems,'' \emph{{IEEE} Trans. Wireless
  Commun.}, vol.~3, no.~4, pp. 1165--1175, Jul. 2004.

\bibitem{Tulino06}
A.~M. Tulino, A.~Lozano, and S.~Verd\'u, ``Capacity-achieving input covariance
  for single-user multi-antenna channels,'' \emph{{IEEE} Trans. Wireless
  Commun.}, vol.~5, no.~3, pp. 662--671, Mar. 2006.

\bibitem{Tse05}
D.~N.~C. Tse and P.~Viswanath, \emph{Fundamentals of Wireless
  Communication}.\hskip 1em plus 0.5em minus 0.4em\relax Cambridge, UK:
  Cambridge University Press, 2005.

\bibitem{Gray06}
R.~M. Gray, \emph{Toeplitz and Circulant Matrices: A Review}.\hskip 1em plus
  0.5em minus 0.4em\relax Now Publishers Inc., 2006.

\bibitem{Smith71}
J.~G. Smith, ``The information capacity of amplitude- and variance-constrained
  scalar {Gaussian} channels,'' \emph{Inform. Contr.}, vol.~18, no.~3, pp.
  203--219, Apr. 1971.

\bibitem{Shamai95}
S.~{Shamai (Shitz)}, ``The capacity of average and peak-power-limited
  quadrature {G}aussian channels,'' \emph{{IEEE} Trans. Inf. Theory}, vol.~41,
  no.~4, pp. 1060--1071, Jul. 1995.

\bibitem{Chan05}
T.~H. Chan, S.~Hranilovic, and F.~R. Kschischang, ``Capacity-achieving
  probability measure for conditionally {Gaussian} channels with bounded
  inputs,'' \emph{{IEEE} Trans. Inf. Theory}, vol.~51, no.~6, pp. 2073--2088,
  Jun. 2005.

\bibitem{Myung06}
H.~G. Myung, J.~Lim, and D.~J. Goodman, ``Single carrier {FDMA} for uplink
  wireless transmission,'' \emph{IEEE Veh. Technol. Mag.}, vol.~1, no.~3, pp.
  30--38, Sep. 2006.

\bibitem{Ochiai12}
H.~Ochiai, ``On instantaneous power distributions of single-carrier {FDMA}
  signals,'' \emph{IEEE Wireless Commun. Lett.}, vol.~1, no.~2, pp. 73--76,
  Apr. 2012.

\bibitem{Li07}
T.~Li and O.~M. Collins, ``A successive decoding strategy for channels with
  memory,'' \emph{{IEEE} Trans. Inf. Theory}, vol.~53, no.~2, pp. 628--646,
  Feb. 2007.

\bibitem{Chan71}
C.~R. Chan, ``Worst interference for coherent binary channel,'' \emph{{IEEE}
  Trans. Inf. Theory}, vol.~17, no.~2, pp. 209--210, Mar. 1971.

\bibitem{Verdu02}
S.~Verd\'u, ``Spectral efficiency in the wideband regime,'' \emph{{IEEE} Trans.
  Inf. Theory}, vol.~48, no.~6, pp. 1319--1343, Jun. 2002.

\bibitem{Varanasi97}
M.~K. Varanasi and T.~Guess, ``Optimum decision feedback multiuser equalization
  with successive decoding achieves the total capacity of the {Gaussian}
  multiple-access channel,'' in \emph{Proc. 31st Asilomar Conf. on Signals,
  Systems and Computers}, vol.~2, Pacific Grove, CA, Nov. 1997, pp. 1405--1409.

\bibitem{Guo02}
D.~G. Guo, S.~Verd\'u, and L.~K. Rasmussen, ``Asymptotic normality of linear
  multiuser receiver outputs,'' \emph{{IEEE} Trans. Inf. Theory}, vol.~48,
  no.~12, pp. 3080--3095, Dec. 2002.

\end{thebibliography}
\end{document}